\documentclass[letterpaper,twocolumn,10pt]{article}
\usepackage{usenix-2020-09}
\usepackage[noadjust]{cite}
\usepackage{amsmath,amssymb,amsfonts}
\usepackage{balance,stfloats,hyperref,soul}
\usepackage[italic]{mathastext}
\usepackage{graphicx}
\usepackage{textcomp}
\usepackage{xcolor}
\usepackage[shortlabels]{enumitem}
\usepackage{tcolorbox}
\usepackage{xspace}

\usepackage{url}
\usepackage{booktabs}
\usepackage[capitalise]{cleveref}
\usepackage{scrextend}
\usepackage{amsmath}
\usepackage{algorithm}
\usepackage{caption}
\usepackage{subcaption}
\usepackage{gensymb}
\usepackage[noend]{algpseudocode}
\usepackage{wrapfig}
\usepackage{hyperref}
\definecolor{mygray}{HTML}{d6d0d6}
\newcommand{\name}{{\tt Eye-Shield}\xspace}
\newcommand{\nameTitle}{\textsc{Eye-Shield}\xspace}
\newcommand{\hidescreen}{{\tt HideScreen}\xspace}
\newcommand{\etal}{\emph{et al.}}
\begin{document}
% \definecolor{background}{HTML}{37474F}
% \pagecolor{background}
% \color{white}

\title{\Large\bf
% \name: Real-Time Shoulder-Surf Protection
%\name: 
%\name: 
\nameTitle: Real-Time Protection of Mobile Device Screen Information from Shoulder Surfing
% \name: Real-Time Shoulder Surfing Protection
}

% \author{\IEEEauthorblockN{\textit{%Anonymous Author(s)
% Submission \# 30,~~~13 pages + references \& appendices}}}

% \author{\textit{Submission \# 30, Anonymous Author(s)}}

\author{
{\rm Brian Jay Tang}\\
University of Michigan\\
{\rm \texttt{bjaytang@umich.edu}}
\and
{\rm Kang G. Shin}\\
University of Michigan\\
{\rm \texttt{kgshin@umich.edu}}
}

% \author{
% \IEEEauthorblockN{1\textsuperscript{st} Brian Tang}
% \IEEEauthorblockA{\textit{Computer Science and Engineering} \\
% \textit{University of Michigan}\\
% Ann Arbor, MI, USA \\
% bjaytang@umich.edu}
% \and
% \IEEEauthorblockN{2\textsuperscript{nd} Kang G. Shin}
% \IEEEauthorblockA{\textit{Computer Science and Engineering} \\
% \textit{University of Michigan}\\
% Ann Arbor, MI, USA \\
% kgshin@umich.edu}
% }

\maketitle

\begin{abstract}
%The usage of mobile devices has become increasingly ubiquitous throughout society. 
People use mobile devices ubiquitously for computing, 
communication, storage, web browsing, and more. 
As a result, the information accessed and stored within 
mobile devices, such as financial and health information, 
text messages, and emails, can often be sensitive. 
Despite this, people frequently use their mobile 
devices in public areas, becoming susceptible to
a simple yet effective attack -- \textit{shoulder surfing}. 
Shoulder surfing occurs when a person near a mobile user 
peeks at the user's mobile device, potentially acquiring 
passcodes, PINs, browsing behavior, or other personal 
information. 
%Since mobile users are only aware 7\% of shoulder surfing incidents, 
We propose,
\name, a solution to prevent shoulder surfers from accessing/stealing 
sensitive on-screen information. \name is designed to protect 
all types of on-screen information {\em in real time}, without
any serious impediment to users' interactions with their 
mobile devices. 
\name generates images that appear readable at close distances, but appear blurry or 
pixelated at farther distances and wider angles. 
It is capable of protecting on-screen information from 
shoulder surfers, operating in real time, 
and being minimally intrusive to the intended users. 
\name protects images and text from shoulder surfers by 
reducing recognition rates to 24.24\% and 15.91\%.
Our implementations of \name achieved 
high frame rates for $1440 \times 3088$ screen resolutions 
(24 FPS for Android and 43 FPS for iOS). \name also incurs 
acceptable memory usage, CPU utilization, and energy overhead. 
Finally, our MTurk and in-person user studies indicate 
that \name protects on-screen information without a large usability cost for privacy-conscious users. 
%A pure software solution, like \name, to shoulder surfing 
%built into devices 
% \name can increase users' awareness and prevent curious or 
% untrustworthy adversaries from accessing %/stealing 
% sensitive on-screen information.
\end{abstract}
% https://ieeexplore.ieee.org/stamp/stamp.jsp?tp=&arnumber=494082
%\begin{IEEEkeywords}
%Privacy, Shoulder Surfing, Mobile Privacy, Real Time, HCI, %Security, Mobile Computing
%\end{IEEEkeywords}

% \newpage

\section{Introduction} \label{sec:introduction}
Mobile devices, such as smartphones, laptops, and tablets, have 
become ubiquitous throughout society~\cite{pew2019mobile}. 
People use them anytime and anywhere to communicate, 
store data, browse content, and improve their lives. 
The information accessed and stored within mobile devices, 
such as financial and health information, text messages, photos,
and emails, is often sensitive and private.

\begin{figure}[t]
  \centering
  \includegraphics[width=\columnwidth]{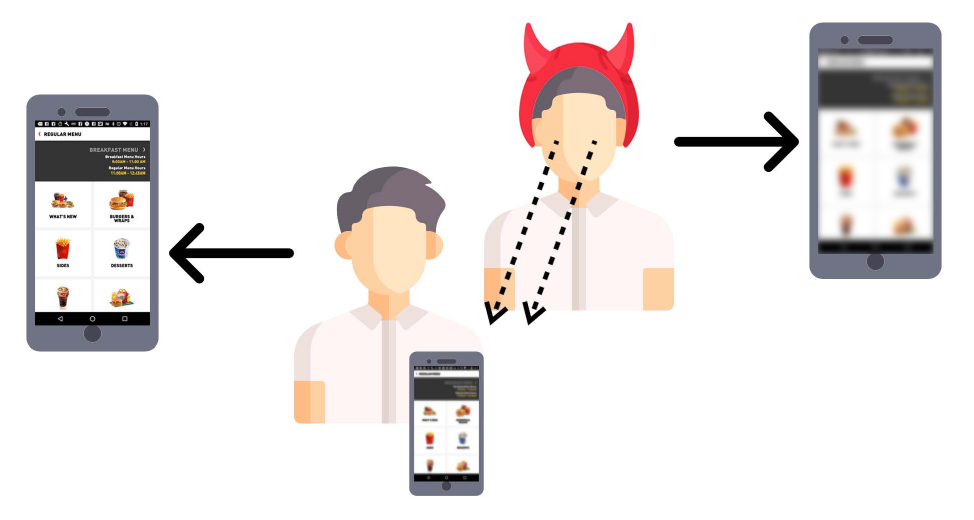}
  \caption{An example scenario of a shoulder-surfing privacy attack that could occur in any public setting. Using \name, the shoulder surfer is prevented from seeing private content on 
  the user's device screen.}
  \label{fig:teaser}
\end{figure}

Despite the private nature of information stored in 
mobile devices, people often choose to use them 
in public areas. 
This leaves users susceptible to a simple yet effective 
attack -- \textit{shoulder surfing}. Shoulder surfing occurs 
when a person near a mobile device user peeks at the user's
screen, potentially acquiring 
sensitive passcodes, PINs, browsing behavior, 
or other personal information. 
This form of visual hacking dates back to the 
1980s when shoulder surfing occurred near public pay phones to 
steal calling card digits~\cite{kee2008Mar}. Shoulder surfing
can be combined with other tools such as cameras or binoculars
to increase the effectiveness of stealing information. 

Studies have shown 
that lack of screen protection in offices leaked information in 91\% of shoulder 
surfing incidents~\cite{globalvisualhacking2016}. 
Another study indicated that 85\% of shoulder surfers 
acknowledged they observed sensitive information they were 
not authorized to see, such as login credentials, personal 
information, contact lists, and financial 
information~\cite{honan2012visualdatasecurity}. 
Experiments indicate 
that you can hack into Snapchat or PayPal accounts by 
peeking at 2-factor authentication codes as 
they appear on a victim's mobile device 
screen~\cite{moore2021snaphack,moore2022paypalhack}.
Shoulder surfing was also found 
to cause negative feelings and induce 
behavior changes~\cite{eiband2017understanding}. 
Some shoulder surfing cases have sparked further discussion. 
For example, in 2017, someone had taken a picture and leaked 
Vitaliano Aguirre II's (Justice Secretary of Philippines) 
smartphone screen during a Senate 
hearing~\cite{aguirre2017furious}, proving he had been plotting against a senator. 
In 2018, Kanye West unlocked his smartphone in front of TV cameras in
the White House revealing that his 6-digit PIN was 000000~\cite{Gartenberg2018Oct}.

%However, most often, shoulder surfing incidents appear in 
%the news whenever automated teller machine (ATM) PIN codes 
%were stolen via shoulder 
%surfing \cite{Kurzweil2015Dec,cbc2019Oct,Times1979Dec}.
Research has also demonstrated that shoulder surfers can 
obtain a 6-digit PIN 10.8\% of the time with just one peek~\cite{aviv2017towards}.
While a person can limit his/her device's susceptibility 
to shoulder surfing by moving to a more private location, 
covering its screen, or turning its display away, these 
measures are not always feasible/effective (e.g., using a 
smartphone on a bus or airplane, using a laptop in an office 
or cafe). These privacy-preserving behaviors are typically 
employed as a response to protect against ``detected" shoulder 
surfers, but studies have shown that mobile device
users are aware of only 7\% of shoulder surfing 
incidents~\cite{eiband2017understanding}. The vast majority of shoulder surfing incidents and information leakage goes unnoticed, making it challenging for users to manually prevent information from being seen by shoulder surfers. Thus, effective defenses either automatically detect and notify 
users of unauthorized shoulder surfers, or continuously
obfuscate information from potential shoulder surfers.
%leaving them unaware of 93\% of shoulder surfing events. 
% Since most shoulder surfing incidents go unnoticed, 

%

% Some information might be time-sensitive or important to view, even in public settings.

Users who seek protection against shoulder surfing 
may wish to hide sensitive information, keep others from 
stealing or peeking at login/PIN credentials, or desire 
peace of mind by having more control over private information. 
Many solutions have been proposed to thwart shoulder 
surfing, but each have their own drawbacks. 
One commonly used privacy-preserving mechanism is a privacy 
film that can be attached to a mobile device 
screen~\cite{GearBrainEditorialTeam2019Aug,3mprivacyfilm}. 
These privacy films only allow light from the mobile device 
display to pass through the film within a narrow viewing 
angle~\cite{android2021Jun}. Users can attach privacy films 
over their smartphone screen to prevent attackers outside of 
a certain viewing angle from seeing any content displayed 
on the smartphone's screen. However, screens covered with 
privacy films are still susceptible to shoulder surfers 
directly behind the user~\cite{Linshang2021Jul}.

% BUILDUP - No one else has developed a general purpose software-based shoulder surfing protection for normal smartphone usage. Solutions are tailor-made to specific apps, messaging services, or are bulky/inconvenient to operate. We develop a method that works on any type of screen content, regardless of color, content type, or frame-rate,

HCI and security researchers have 
explored various other defenses against shoulder surfing. 
They can be categorized into three main screen protection 
types: 1) shoulder surfer detection, 2) software solutions, 
and 3) authentication-specific approaches. 
Each of these solutions has its own advantages 
and drawbacks which we will discuss thoroughly in \cref{sec:background}. 
% For example, some solutions have been tailor-made to specific apps~\cite{von2016you} 
% or messaging services~\cite{eiband2016my}, or are inconvenient and disrupt 
% the user's desired/intended activity~\cite{blackberry2022privacyshade}.
No software-based defense has been 
developed for protecting the real-time usage of mobile devices, 
such as watching videos, playing games, and interacting with UI animations. Prior solutions are 
neither comprehensive nor capable of protecting all types of information from leaking to shoulder surfers.

Our goal is to prevent the leakage of 
\textit{all} sensitive on-screen information to shoulder 
surfers without interrupting the intended user's 
device usage. To address this challenge, we develop \name, a software solution 
that protects any on-screen information from shoulder surfers 
{\em in real time}. \name can protect colored images, text, 
mobile app UIs, videos, and smartphone browsing from 
shoulder surfers. In this sense, it is a software version 
of privacy film protecting against shoulder surfers
from any angle. While it protects information from shoulder surfers, it
simultaneously allows the intended users
to still view and comprehend the on-screen content. We envision that \name can be deployed 
either as a feature of the device's operating system or an API 
for apps. Having a software solution to 
shoulder surfing implemented across mobile device platforms can increase 
awareness and protect sensitive information.

Throughout the development of \name, we identify 
and address the following requirements/challenges:
\begin{enumerate}[noitemsep,leftmargin=0.4cm,topsep=5pt]
    \item The protection of \textit{any} type of information 
    displayed on-screen from shoulder surfers using purely software has not been 
    developed before. Prior work has been
    tailored to protect certain types of information from
    shoulder surfers. In contrast, we explore and develop a universal shoulder
    surfing defense mechanism that functions with {\em any} type of 
    content displayed on a device's screen. 
    \Cref{sec:design} details how the design of \name achieves this. \name protects images and text from shoulder surfers by reducing recognition rates
    to 24.24\% and 15.91\%. Sections \ref{sec:methodology} and \ref{sec:results} present an extensive evaluation of \name's protection guarantees using several datasets of images, user interfaces (UIs), and videos.
    
    \item Protecting on-screen information from shoulder 
    surfers in {\em real time} without interruption to the 
    intended device usage has not yet been addressed by purely 
    software-based defense mechanisms. Previous systems rely on pre-rendering images 
    and texts, or require the user to manually obfuscate 
    portions of the screen. In contrast, \name protects
    the entire screen with minimal to no disruptions to the 
    intended device usage, while meeting real-time 
    constraints. The implementation of \name in mobile devices while achieving
    real-time performance, e.g., at a rate of 43 FPS on iOS devices for even the largest screen 
    resolution sizes. 
    \name can achieve smoother performances of 60+ FPS by 
    reducing the screen resolution.
    \Cref{subsec:results_performance} provides performance details, 
    such as latency, memory usage, CPU utilization, and energy overhead.

    \item \name must achieve the same level of protection that 
    a privacy film provides, even if the shoulder surfer peeks 
    at the screen from directly behind the intended user. Additionally, 
    \name should not cause noticeable usability
    disruptions, similar to a privacy film.  \Cref{subsec:method_mturk_study,subsec:method_user_study,subsec:results_mturk,subsec:results_usability} detail the usability of \name with two user studies. The 
    first study (with \textit{n} = 22 participants) demonstrates that users 
    can still understand and use a mobile device employing \name 
    while shoulder surfers are unable to comprehend the screen 
    content. The second study (\textit{n} = 99) scales the
    findings of the first study. The participants found \name 
    to be easily usable, and privacy-conscious participants were 
    satisfied with \name's protection and its quality degradation.
\end{enumerate}

% \newpage

\section{Background} \label{sec:background}

\subsection{Threat Model} \label{subsec:threat_model}
\begin{figure}[t]
  \centering
  \includegraphics[width=0.5\columnwidth]{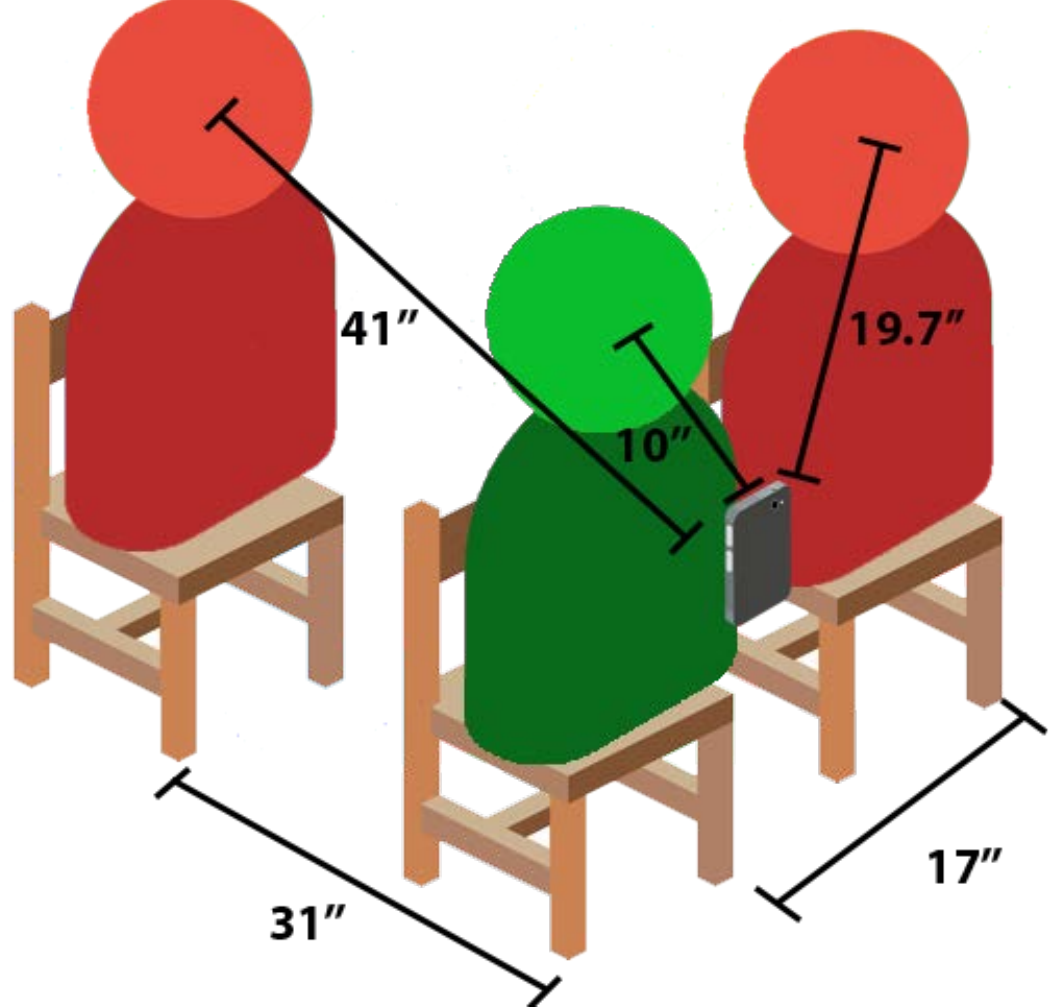}
  \caption{A diagram depicting the shoulder surfing threat model we consider. The intended user in green must be able to see and comprehend the information displayed on a screen.}
  \label{fig:threat_model}
\end{figure}

A typical shoulder surfing adversary is a curious or 
malicious person who seeks to observe or steal the 
information displayed on a victim's device. 
Most shoulder surfers do not wish to get caught, and hence
we assume they would gather information stealthily,
i.e., the adversary peeks at the victim's screen either 
from an angle or a distance behind the victim. 
In realistic scenarios, an adversary who can view 
information on a victim's device is sitting 
either behind or next to the victim (e.g., public 
transportation seating, cafe, restaurant, 
auditorium, lecture, or office settings). 
In an airline setting, often the most cramped seating arrangement, 
the standard seat pitch is 31'', and the standard 
seat width is 17''~\cite{airline2018Apr}. 
The majority of smartphone screen sizes 
are between 5'' and 7''~\cite{statista2022Mar}, 
with a resolution between 720$\times$1280 and 
1440$\times$2960~\cite{deviceatlas2019Oct}, 
with most users operating them at a distance 
of 10''~\cite{yoshimura2017smartphone}. 
Our system should thus protect on-screen information from 
adversaries 41'' directly behind the user's screen and 20''
beside the user's screen. 
\Cref{fig:threat_model} depicts this threat model. 
We use these measurements as the baseline setting and 
threat model for our experimental evaluation.
% Likewise, laptop screens can range from 11'' to 17.3'', a resolution 
% ranging from 1280$\times$720 and 3840$\times$2160, with most users 
% operating them from 20'' away~\cite{Granquist2018Sep}. 
 
We assume that the adversary can either peek at the screen 
or use a camera to record the content on the victim's 
device. Since most shoulder surfing incidents are known to be out of curiosity rather than with malicious intent, 
\name is not designed with the intent of protecting victims from 
adversaries using highly sophisticated tools or attacks. This, in turn, keeps the deployment of \name easy and inexpensive.
We also assume the adversary is interested in \textit{any} type 
of content displayed on the victim's screen, 
not just passwords, text, or PIN entry. 

\subsection{Shoulder Surfing Defenses} 
\label{subsec:related_work}
Several existing defenses against shoulder surfing 
have been proposed. These defenses 
can be categorized into 4 main types: privacy films,
authentication mechanisms, shoulder surfer detection,
and software solutions.

\noindent
\textbf{Privacy Films:}
The most ubiquitous privacy defense against shoulder 
surfing is the privacy film. The film can be attached 
to a smartphone or laptop screen
\cite{GearBrainEditorialTeam2019Aug,3mprivacyfilm}, and only allows 
light from the mobile device display to pass through the 
film within a narrow viewing angle~\cite{android2021Jun}. 
Users can attach privacy films on their smartphone screen
to prevent attackers outside of a certain viewing angle 
from viewing any content presented on the smartphone's 
screen. However, screens covered with privacy films are 
still susceptible to shoulder surfers directly behind the
user. Additionally, purchasing a privacy film incurs an 
added cost and must be physically attached to the 
mobile device screen -- a price not all users are willing 
to pay.

\noindent
\textbf{Authentication Mechanisms:}
The problem of shoulder surfing has been most extensively 
explored in the context of authentication, particularly 
for PIN entry, password entry, game-based authentication, 
and drawing patterns for authentication. IllusionPIN uses 
hybrid images which encode the low spatial frequencies of 
one PIN digit with the high spatial frequencies of a 
different PIN digit to make the PIN entry appear 
different to a far-away shoulder 
surfer~\cite{papadopoulos2017illusionpin}. Zakaria \etal 
\cite{zakaria2011shoulder} implemented authentication by 
having users draw secret patterns. Kumar \etal 
\cite{kumar2007reducing} developed a system that allows 
users to enter passwords using just eye gaze on a keyboard. 
Another gaze-based authentication mechanism developed by Abdrabou 
\etal~\cite{abdrabou2019just} used a combination 
of eye gaze and hand gestures as a potential mechanism 
for authentication. 

\noindent
\textbf{Detection of Shoulder Surfers:}
This approach prevents shoulder surfers from 
unauthorized viewing of mobile device screens by 
detecting an additional set of eyes focusing on the 
screen and alerting the user to the potential shoulder 
surfing activity~\cite{bradshaw2021chromebooks,lian2013smart}. 
Upon detecting a shoulder surfer, the system can either 
issue an alert and pause user activity, or the system 
could apply another software-based shoulder surfing 
protection mechanism.

\noindent
\textbf{Software Solutions:}
Software-based shoulder surfing defenses do not require any 
additional hardware beyond the typical functionalities available 
on a mobile device. The simplest of software implementations 
protect screen privacy by gray-scaling or darkening the 
screen. These approaches can also limit the visible regions 
of the screen via selective hiding and selective showing of 
screen content. There have been several approaches proposed by 
Zhou \etal~\cite{zhou2015somebody}, Khamis \etal~\cite{khamis2018eyespot}, 
and BlackBerry~\cite{blackberry2022privacyshade}.
Notably, the solution developed by Khamis
uses eye-tracking to automatically selectively display and 
hide regions of the screen. Other software-based defenses 
aim to obfuscate specific types of information such as text. 
Eiband \etal~\cite{eiband2016my} allow users to use their 
handwriting as the font, to increase the reading 
difficulty for shoulder surfers while maintaining font 
familiarity with the intended user. 
Von Zezschwitz \etal~\cite{von2016you}~seek to 
protect photo gallery browsing from shoulder surfers 
by pixelating or crystallizing the displayed images. 
This way, adversaries unfamiliar with the photos will be 
unable to extract meaningful information from peeking at a 
victim's device. Finally, \hidescreen uses a grid-based 
approach to make the low spatial frequency components of an 
image or text appear like a completely gray background to 
shoulder surfers~\cite{chen2019keep}. \hidescreen encodes 
information as high spatial frequency components visible 
only to viewers close to the screen.

\noindent
\textbf{Other Related Approaches:}
Apple proposed using AR glasses to un-blur mobile device 
screen content, such that it appears normal to the glasses 
wearer, by adjusting the prescription of the smart glasses 
for the wearer~\cite{patent2021apple}. However, this 
approach has yet to be implemented, and it relies on 
additional expensive hardware not readily available to 
all users.

% PrivacyScout designs a systematic framework 
% for evaluating the shoulder surfing risk of mobile 
% devices~\cite{bace2022privacyscout}.

% \newpage

\section{Design, Implementation, and Comparison}  \label{sec:design}
This section details \name's design and implementation, and
compares it with related work.
\subsection{Design}

\name is designed to present the original screen content with only 
minor quality degradation to the intended user, but it can render shoulder surfers 
beyond a certain distance/angle away from the screen unable to discern 
the screen content. \name achieves this by leveraging the fact that 
at a sufficient distance, it is impossible for an optical system to 
distinguish between two nearby light sources. By applying this theory 
of resolving power~\cite{singh2015fundamentals}, we can construct checkered grids of pixels 
that can appear individually discernable at a close distance, 
but appear as a uniform average of the projected colors.

% \input{FiguresTex/fig_color_example}

% as follows: 1) an image of a smartphone screen, photo, video, or app UI is input into the system, 2) a checkered grid mask is computed with the same dimensions as the input, 3) the input image is blurred or pixelated and used as the target, 4) the \name algorithm uses the original input, grid mask, and target to compute a shoulder surfing resistant image, and 5) the protected image is output from the system.

\begin{figure}[t]
  \centering
  \includegraphics[width=0.8\columnwidth]{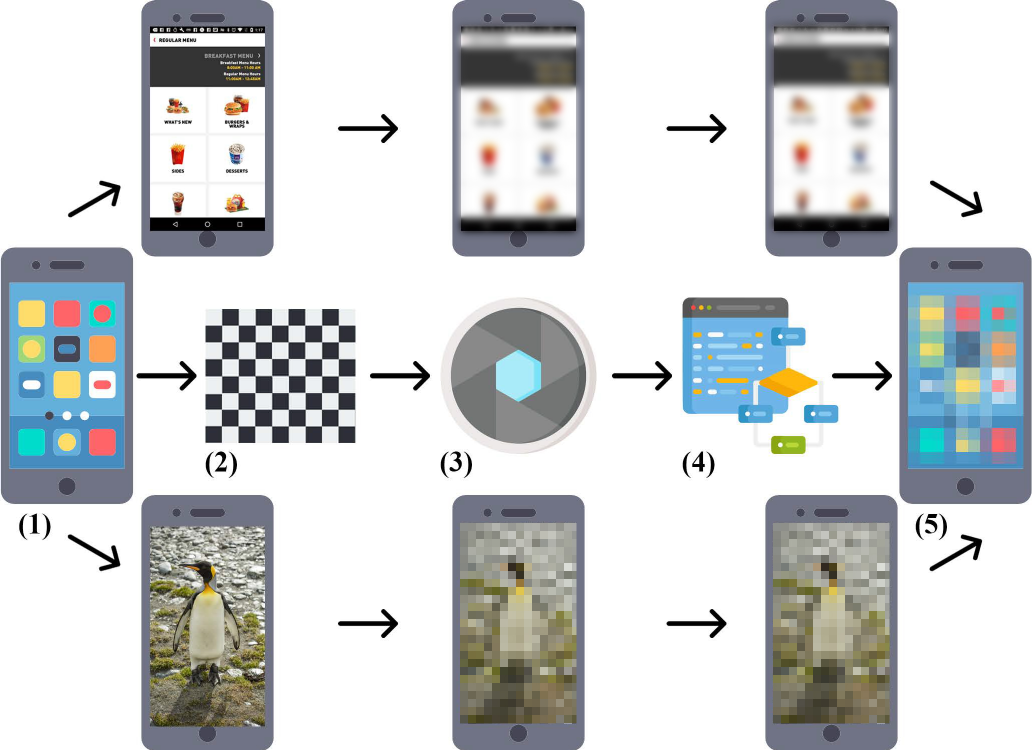}
  \caption{\name works as follows: 1) an image or screen is input into the system, 2) a checkered grid mask is computed with the same dimensions as the input, 3) the input is blurred/pixelated, 4) \cref{alg:complement} uses the input, grid, and blurred input to compute, 5) the protected image.}
  \label{fig:pipeline}
\end{figure}

\cref{fig:pipeline} provides an overview of the processing required 
to protect on-screen information.
The main components required for the \name algorithm are 1) the original screen/image, 
2) a checkered grid mask in the dimensions of the original image, and 
3) a blurred or pixelated version of the original image. 
Then, the protected output image will be computed with \cref{alg:complement}.

\noindent
\textbf{Grid Generation:}
\name's design is based on the observation that at angles smaller 
than $1.22\lambda/D$, where $\lambda$ is the wavelength of light,
and $D$ is the lens aperture, it is no longer possible to 
distinguish two light sources from one another. So, using a 
checkered grid of pixels results in the pixels 
appearing as one uniform color to a user viewing from a far 
distance, whereas a user near the screen can 
distinguish between the individual pixels within the grid. 

\noindent
\textbf{Blurring/Pixelation:}
To enable \name to function on colored content of any type, 
our system design makes the screen appear blurry (for text and 
mobile UIs) or pixelated (for images and videos) to users who perceive
the pixels on the device screen from a small resolving power angular resolution (around 30--40'' away from a smartphone
or 20'' with a 45\degree angle). 

\noindent
\textbf{Computing Average Colors:}
We observe that two colors arranged in a checkered grid pattern and 
displayed on a screen appear as their average (additive color). 
While the best perceptual approximation of this averaged color can be 
achieved using a color approximation model such as
CIECAM02~\cite{moroney2002ciecam02} or CIELAB~\cite{zhang1996spatial}, 
due to real-time computation constraints, we implement this color 
averaging as the root mean square in \cref{eq:rms} to reduce the required
computation time. Using the target (blurred/pixelated) image pixels 
as $rms$ and the original image pixels as $x$, \cref{eq:rms}
computes the protected output image pixels as $y$.

Overall, after acquiring the required original image, target, 
and grid, to compute the protected image, we can represent
\cref{eq:rms} using the pseudocode in \cref{alg:complement}:

\begin{algorithm}[!b]
\small
\label{algorithm}
\caption{Where $\mathsf{img}$ is the original $w \times h \times 3$ image\\
where $\mathsf{grid}$ is a $w \times h$ checkered grid of 1s and 0s\\
where $\mathsf{targ}$ is the $w \times h \times 3$ image, blurred or pixelated}
\begin{algorithmic}[1] \label{alg:complement}
\Procedure{\name algorithm}{$\mathsf{img}, \mathsf{grid}, \mathsf{targ}$}
\State $\mathsf{complement} = \left(\mathsf{targ}^{2} \cdot 2\right) - \mathsf{img}^{2}$
\State $\mathsf{delta} = \left(\mathsf{complement} - \mathsf{img}^{2}\right) \cdot \mathsf{grid}$
\State $\mathsf{newimg} = \sqrt{\mathsf{img}^{2} + \mathsf{delta}}$
\State $clip\left(\mathsf{newimg}, 0, 255\right)$
\EndProcedure
\end{algorithmic}
\end{algorithm}

\begin{figure}[t]
  \centering
  \includegraphics[width=0.8\columnwidth]{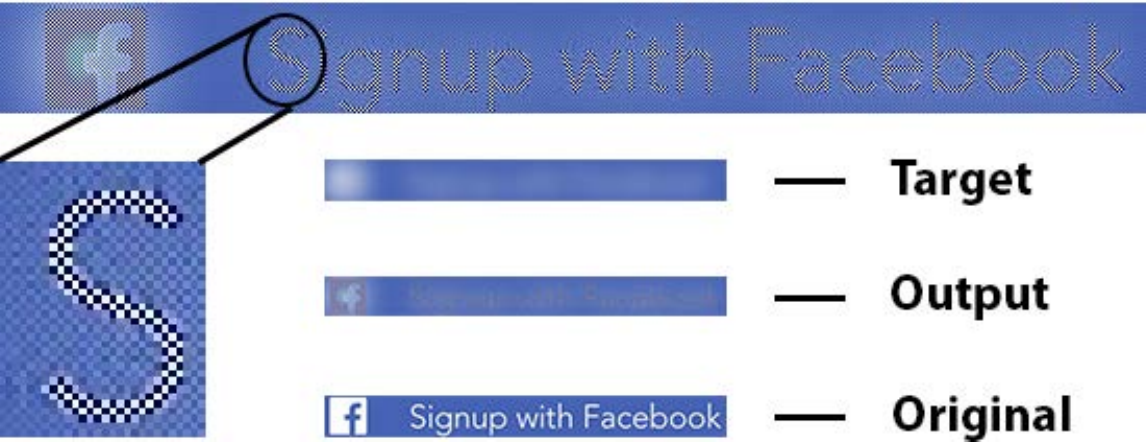}
  \caption{A close up look at a protected line of text.}
  \label{fig:closeup}
\end{figure}

An example of the resulting protected screen content can be seen in
\cref{fig:closeup}, where a full-size version of the protected content
and downscaled versions of the target, protected output, and original
text are presented. Note that the text output by \name is much harder to discern compared to the original line of text. However, viewing the text in close proximity allows the user to still read the content.

\subsection{Implementation of \nameTitle}

\name utilizes the device's GPU on 4 platforms (Windows, Android, 
MacOS, iOS) to accelerate image processing and matrix operations. 
In the Windows implementation, we created both a CPU-only version and a
GPU version for performance comparisons. The desktop implementation
also supports video processing and writing using FFMPEG~\cite{tomar2006converting}. 
The full development stack of \name for each platform can be 
found in \cref{table:stack} in the appendix.
% \footnote{We plan to release our code at the following URL: \url{https://osf.io/haxzb/?view_only=8629a44c79d74b71b603be3830f60583}.}
As it is unrealistic to expect app developers
to implement a shoulder surfing solution on their platforms, 
\name was devised as a proof-of-concept solution to be implemented
on mobile device operating systems. As such, \name acts more like 
a screen filter than an API for mobile app developers.

% Include that I created shaders/custom kernels for each of these platforms

% Include specific caveats, details with each implementation
% Include thread count, thread grouping, other parameters, etc.
\noindent
\textbf{PC, CUDA:}
On several workstations and servers with access to an Nvidia GPU and 
CUDA drivers, \name is able to run in real time using Python, OpenCV, 
and CUDA. \name leverages CUDA to perform image blurring, grid 
generation, and the matrix operations used to compute the 
average colors.

\noindent
\textbf{Android, Vulkan:}
\name runs in real time on Android mobile devices using C++, OpenCV, 
and Vulkan. \name is capable of achieving real-time performance in 
image blurring without using OpenCL drivers. 
% It was unable to access 
% the OpenCL drivers on the Samsung Galaxy S20 device used in 
% development, due to restrictions placed on OEM unlocking for newer 
% Android versions and Qualcomm Snapdragon devices without SIM cards. 
We note that \name's implementation could improve with access to 
OpenCL drivers. \name leverages Vulkan to perform the matrix 
operations used to compute the average colors. Here, the 
implementation of the grid generation is quick enough to be performed 
on the CPU in real time.

\noindent
\textbf{MacOS, iOS, Metal:}
\name can run in real time on MacOS and iOS devices using Swift, 
CoreImage, and Metal. \name leverages Metal to perform image blurring 
and the matrix operations used to compute the average colors. 
Grid generation is performed in C++ on the CPU.

\subsection{Comparison with Related Work}
\label{subsec:comparison}

\Cref{table:relatedwork} in the Appendix contains a comprehensive checklist of some
notable shoulder
surfing defense mechanisms and the types of content they can protect (not detection-based).
\name is capable of protecting all information categories except for 
PIN entry, since keypad reshuffling is required to make PIN entry fully secure.

\noindent
\textbf{Comparison to \hidescreen~\cite{chen2019keep}:}
The design of \name is inspired by 
\hidescreen~\cite{chen2019keep}, which leverages an 
observation that the human optical system is incapable of 
distinguishing between two adjacent light sources beyond 
a certain angle. This limit of the resolving power of an 
optical system can be leveraged to hide information from 
shoulder surfers on mobile devices by using a grid pattern 
of light and dark components so that it appears uniformly 
gray from far distances. \hidescreen's design suffers 
from several key limitations: 1) selection of dark 
and light components limits the design to gray-scale images 
and text, 2) a significant amount of information is 
lost by only using 6 different grid patterns, and 3) the 
latency of hiding a $512\times512$ image is 1684ms, 
a run-time unsuitable for real-time screen usage.

\name utilizes a similar grid-based design to shield 
on-screen content from shoulder surfers. However, the 
introduction of 3 channels for color poses a new challenge: 
the ability of \name to preserve color information on 
screens. The approach of 
averaging everything into a monotone gray color fails to 
protect information from shoulder surfers when applied to 
full-color images. \name does not attempt to perfectly hide all screen 
information from shoulder surfers, but it overlays 
screen information with a grid to make the screen appear 
blurry or pixelated to a far-away shoulder surfer. 
Besides enabling the protection of all types of screen content, \name also achieves the performance required for real-time usage. These features allow for its integration with prevalent mobile device OSs and apps to protect a broad range of colored content, videos, and real-time browsing. It also removes many undesirable requirements in \hidescreen such as needing app developers to adopt the \hidescreen platform or display calibration, a time-consuming process. Finally, \name causes no interruptions to usage and changes significantly less content than \hidescreen. Overall, \name's performance and usability is significantly improved over \hidescreen.

\noindent
\textbf{Comparison to Privacy Films~\cite{3mprivacyfilm}:}
In many ways, \name is most similar to privacy films in terms of
functionality and interface. \name is designed to protect and redraw
the entire screen with minimal impacts on the screen refresh rate. 
The main differences are that 
privacy films are unable to protect against shoulder surfers directly 
behind the user, privacy films completely darken the screen from a 
shoulder surfer's perspective, and privacy films require users to be
aware of shoulder surfing risks and be willing to pay and install films across all their devices. \name protects against {\em any} shoulder surfers both beside and 
behind users, blurs and pixelates the screen rather than darkening it, 
and is free and implemented entirely in software.

\noindent
\textbf{Comparison to Zezschwitz \etal~\cite{von2016you}:}
A photo browsing software capable of protecting information from shoulder surfers was developed 
by Zezschwitz \etal~\cite{von2016you}. It provides a somewhat similar
implementation to \name in its usage of blurring, pixelation, and 
crystallization of images within the gallery. However, using the 
generated grid, \name allows the intended user to still comprehend 
information and view low-level details within the pixelated images. The system %by Zezschwitz \etal
in \cite{von2016you} completely blurs and pixelates content, even for the intended user.
% Each of these implementations of shoulder surfing protection 
% mechanisms requires parameters to be tuned for each task
% and app developers to adopt these techniques.

\noindent
\textbf{Comparison to Eiband \etal~\cite{eiband2016my}:}
This implementation %of Eiband \etal~\cite{eiband2016my} 
allows users to view text messages in their personal handwriting. 
This personalization makes it more difficult for shoulder surfers to read
information on the user's device. \name causes text to appear blurry
to shoulder surfers without the need for customization or
personalization. This allows users to more quickly
adopt the privacy mechanism, further reducing barriers to usage.

\noindent
\textbf{Comparison to Blackberry Privacy Shade~\cite{blackberry2022privacyshade}:}
This software provides users with a tool for
darkening all portions of the screen except for a small section.
Users can then closely protect and hide the information from nearby
shoulder surfers using their hands and body. This requires additional
effort on the user's part and can interrupt a user's typical task
flow. The implementation of \name allows users to use their devices
without actively worrying about their information by
protecting the entire screen.

\noindent
\textbf{Comparison to IllusionPIN~\cite{papadopoulos2017illusionpin}:}
IllusionPIN uses hybrid images and keypad shuffling to protect PIN entry
from the gaze of shoulder surfers. The combination of both tools makes the
PIN appear to faraway shoulder surfers as a normal keypad, but the intended
user can see the actual arrangement of the PIN numbers. While \name does not 
explicitly support the protection of keypad entry, the combination of keypad 
shuffling and blurring can cause shoulder surfers to have difficulty in 
reading a user's PIN information.

\begin{figure}[t]
  \centering
  \includegraphics[width=0.9\columnwidth]{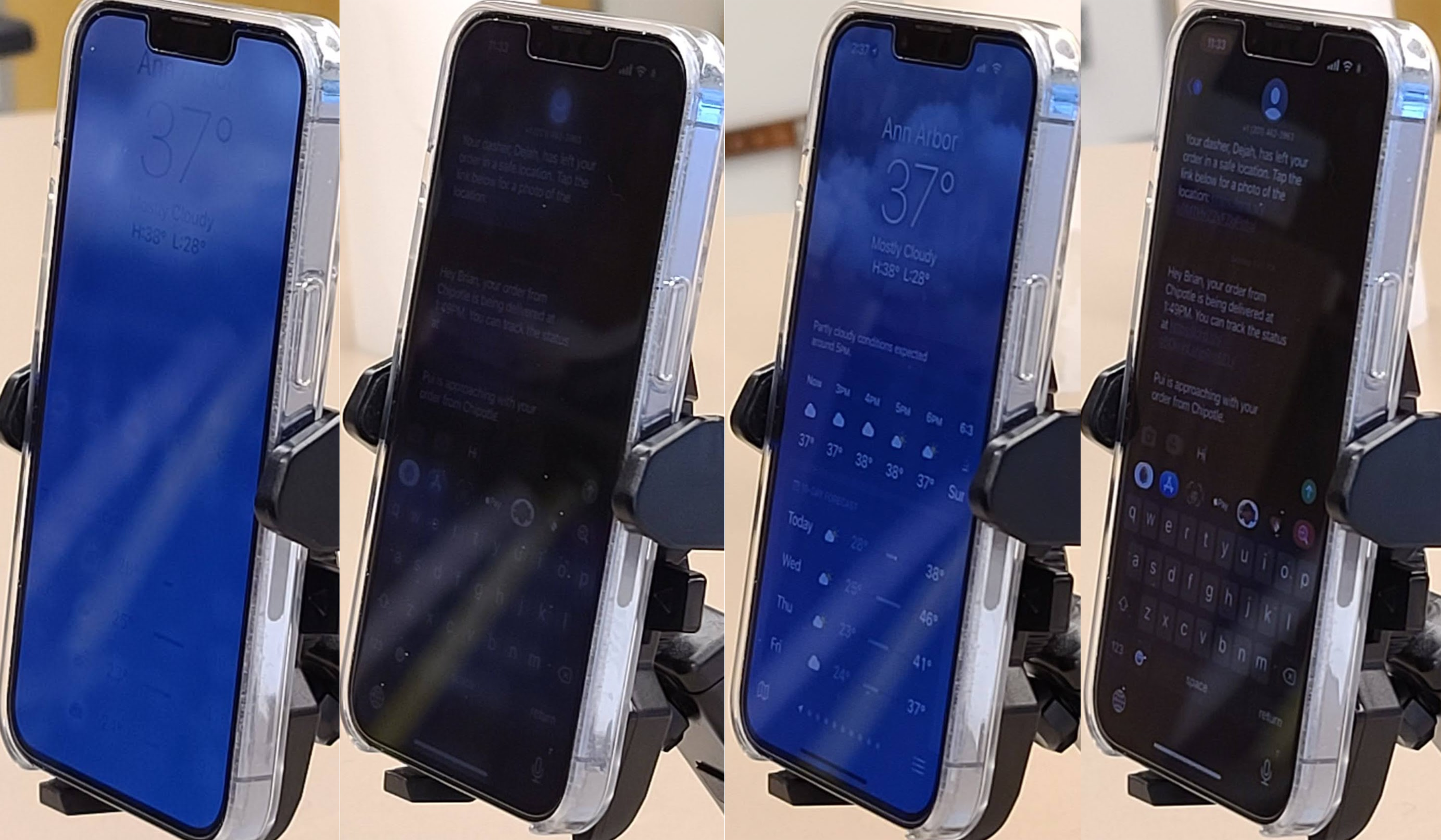}
  \caption{Photos of screens protected by \name (left) and the unprotected original screens (right). Photos were taken at a distance of 19.7'' and an angle of 45\degree, from the perspective of a shoulder surfer. The images were captured with a 108MP, f/1.8, 26mm smartphone camera at $3\times$ zoom.}
  \label{fig:example_angle_zoom}
\end{figure}

% \newpage

\section{Methodology for Evaluation of \nameTitle}
\label{sec:methodology}

\noindent We empirically evaluate \name's efficacy in protecting 
content, performance and resource consumption, and usability cost. 
This section describes the experimental setup and design details for 
each evaluation criterion as follows:
\begin{itemize}[noitemsep,leftmargin=0.4cm,topsep=5pt]
\item \textbf{Protection: } \cref{subsec:method_perceptual_similarity,subsec:method_semantic_performance,subsec:method_mturk_study,subsec:method_user_study}
\item \textbf{Performance: } \cref{subsec:method_performance_evaluation}
\item \textbf{Usability: } \cref{subsec:method_mturk_study,subsec:method_user_study,subsec:method_interview}
\end{itemize}
Our evaluation results will be provided in Section~\ref{sec:results}.

\subsection{Perceptual Similarity} \label{subsec:method_perceptual_similarity}

We generate protected images and videos using \name on 3 datasets: 
1) RICO, an image dataset of mobile app UIs~\cite{Deka:2017:Rico},
DIV2K, a diverse dataset of high resolution (2K) images
\cite{Agustsson_2017_CVPR_Workshops}, and DAVIS, a video dataset 
for object segmentation~\cite{caelles20192019}. 
With the combined datasets, we evaluate 3,882 unique images. 
Using 4 different parameters for grid size and blurring/pixelation 
intensity, we generate a total of 124,224 protected images 
for use in our experimentation. 
To evaluate the information protection guaranteed by 
protecting images with \name, we measure the SSIM 
index~\cite{wang2004image} (structural similarity) of the 
protected image and the blurred/pixelated target image. 
SSIM extracts and compares the luminance, contrast, and 
structure between two images. 
The formula is given in \cref{eq:ssim}, where $x,y$ are windows 
of size $N \times N$, $\mu$ is the average, $\sigma$ is the 
variance/covariance, and $c$ are normalization constants. 
In our experimentation, we use SSIM with $x,y = 7 \times 7$, 
and $c_1 = 0.01, c_2 = 0.03$. To simulate the distance from 
which a shoulder surfer views the protected image, we downscale 
the protected image and compare the SSIM with the target 
(blurred/pixelated) image.
We also measure the SSIM of each full-scale protected image 
compared with the original image to evaluate the extent to which 
the protected image represents the original image.

% ~\cite{wang2004image}
% ~\cite{zhang2018unreasonable}

\subsection{Semantic Performance} \label{subsec:method_semantic_performance}

% Google cloud details
To determine whether text and high-level details from images 
can be hidden from shoulder surfers by \name at a large-scale, 
we leverage the Google Cloud Vision API to perform image 
recognition and optical character recognition (OCR). 
The ``label detection'' (image recognition) service provides 
labels and their associated confidence scores. Using OCR, 
we also detect the boundaries of texts and extract 
their content from images of mobile UIs. We evaluate the efficacy 
of \name by performing label detection on downscaled protected 
images/videos from the DIV2K and DAVIS datasets and performing 
OCR on the mobile app UI screenshots from the 
RICO dataset. These results are compared with the API outputs of 
the unprotected downscaled images to provide a baseline for 
the percentage of labels and text protected.

\subsection{Performance Evaluation} 
\label{subsec:method_performance_evaluation}

To determine whether \name can run in real time, we benchmark the 
processing time and memory consumption on various devices, both with 
and without leveraging the devices' GPUs. We test a wide range of 
image resolutions --- as small as $256 \times 144$ and as large as 
$3088 \times 1440$ (see \cref{table:resolutions}). 
These encompass commonly-used video resolutions, mobile screen 
resolutions, and an image size for direct performance comparisons 
with \hidescreen. The overall processing time of \name is derived 
from a combination of the grid generation, blurring/pixelation of 
the original image, and the screen hiding
algorithm that computes complementary colors.
The performance data are gathered by logging 
processing times after running \name for 100 image frames. 
We also evaluate the resource overhead of \name by recording 
the maximum CPU utilization and the maximum memory usage after 
running \name over the stream of 100 images. 
Finally, we measure energy consumption by using the Android 
Studio energy profiler~\cite{energyandroid2020Aug} and 
the Xcode energy impact gauge~\cite{energyios2016Sep}.
These energy impacts are estimated based on GPU and CPU utilization,
network and sensor usage, as well as other costs/overheads. 
The performance evaluations were run on 4 devices: a workstation 
with an AMD Ryzen 9 3900X CPU and an Nvidia GTX 2080 Super GPU, 
a 2021 MacBook Air with an M1 chip, a Samsung Galaxy S20 
Ultra, and an iPhone 13 Pro. 
See \Cref{table:devices} for more details.

\subsection{MTurk Study} \label{subsec:method_mturk_study}

We conducted a user study through an Amazon Mechanical Turk (MTurk) 
survey to assess the protection strength of \name on a diverse 
set of images and videos. 99 U.S. participants, aged 23--71 
($M=43.19$, $SD=10.37$; 55\% men, 44\% women), completed 
our survey. Our user study protocol was exempted (and approved) 
by our institution IRB, and participants 
who completed the study received \$1.75 as compensation. 
We developed a series of questions where participants are presented 
with the original images/videos and the images/videos protected by 
\name (in random order). To mitigate bias, participants were shown the protected images from the shoulder surfer's perspective, followed by the protected images from the intended user's perspective, and finishing with the unprotected images from the shoulder surfer's perspective.  Participants are asked several text entry
questions regarding the content within each image. 
To represent the distance at which a shoulder surfer sees the 
content, we also present a ($4 \times$) down-scaled version of 
both the original and protected content. 

To derive the $4 \times$ downscaling, we calculate the angular 
diameter of an 5.78'' iPhone 13 Pro (the device used in 
our in-person user study) at 2 distances, 10'' and 41'' 
(10'' + 31'', the average airplane seat pitch). 
We obtain an angular diameter 
of 8.064\degree and 32.239\degree, or roughly a $4 \times$ 
perceived size difference. 
% Angular diameter is also referred to as {\em apparent size} and is 
% used to calculate how big an object appears from a certain distance. 
% Eq.~(\ref{eq:angular_size}) contains the equation for calculating the 
% angular diameter ($\delta$) of a device with a screen size of $d$, 
% a distance of $D$ from the user's eyes.

We collected responses from participants perceiving the 
protected and unprotected screens from the perspectives 
of the shoulder surfer and the intended user. 
We presented a total of 20 unique images, 6 unique videos, and 20 
unique mobile app UIs to participants. These images and videos
were randomly sampled from our evaluation datasets. 
Each participant answered questions regarding a random subset 
of 8 unique images/videos, portrayed as the downscaled protected 
screen, the full-size protected screen, and the downscaled 
original screen (24 images/videos per participant). 
Our survey averaged around 12.62 minutes for completion. 
With 99 participants, each question received an average of 
19.8 responses, for a grand total of 3,180 responses. 
We measure a shoulder surfer or intended user's recognition 
rate (binary accuracy, $R_{ss}$, $R_{iu}$), or the percentage 
of text, images, and videos correctly labeled by the
MTurk participants. 
% Using the responses for the
% original unprotected images as a ground truth, we can obtain \name's
% protection rate ($R_{iu} - R_{ss}$). 
In addition to evaluating the efficacy of \name through this 
study, we also obtain user perceptions towards shoulder surfing 
and the users' inclination to use the protected screens. 
These responses were obtained using 5-point Likert survey 
responses ranging from ``Strongly Disagree'' to ``Strongly Agree'',
normalized to values between 0--4. 
Finally, we recorded the response time for each question to 
better understand how \name impacts the comprehension 
time of protected images. Sections \ref{subsec:results_mturk} 
and \ref{subsec:results_usability} provide the MTurk study results.

\subsection{User Study} 
\label{subsec:method_user_study}

We conducted an additional in-person user study to assess the 
usability and protection strength of \name on a diverse set 
of images and videos.
22 U.S. participants, aged 22-63 ($M=36.32$, $SD=13.15$; 
41\% men, 59\% women), completed our user study. We recruited participants with varying degrees of smartphone experience and visual health.
Our protocol for this in-person user study was approved and 
exempted by our University IRB. 
The user study was conducted in a brightly lit lab with the 
device brightness at a moderate setting. We ensured the 
participants' vision was unobstructed by glare before 
continuing further in the study. Prior to conducting the user 
study, we placed the device (iPhone 13 Pro) displaying the 
images and videos protected with \name on a smartphone mount 
on a table. The screen brightness of the device was set to 66\%. Participants were instructed not to move or 
alter the device. This was done to ensure consistency between the evaluated study settings and to avoid the additional confounding factors that would arise if participants held the device. They were asked to evaluate the presented 
screen in 5 different settings: 1) a shoulder surfer 41'' 
away from the screen protected with \name, 2) an intended user 
10'' away from the screen, 3) a shoulder surfer 20'' away from the protected screen at a $45 \degree$ angle, 
4) a shoulder surfer 41'' away from the unprotected screen, and 
5) a shoulder surfer 20'' away from the unprotected screen at a $45 \degree$ angle. 
This order was selected to mitigate bias from previous tasks. 
We also asked participants to evaluate a screen protected by a privacy film compared to 
a screen protected by both \name and a privacy film by asking them to lean towards the device 
in setting 3 until they were able to read the content displayed on the device screen. 
We approximated the distance they had to lean. 
Setting 3, setting 5, and the evaluation with the 
physical privacy film were changes made to the original user 
study after suggestions from reviewers. This expanded user 
study along with the qualitative interview 
(\cref{subsec:method_interview}) was conducted with 15 out 
of the 22 total participants. The total study took around 
1 hour to complete with \$20 for compensation.
We developed a series of questions where participants are 
presented with the original image/video and the image/video 
protected by \name (in random order). We presented a total 
of 6 images, 2 videos, 7 mobile app UIs, and 2 screen 
recordings to participants. 
Participants were asked several questions regarding the 
content within each image, involving reading text, describing
images, and explaining videos. Some examples of these tasks 
were questions like ``What is the current card balance?'', 
``Can you read the first word in each sentence?'', and 
``Can you describe the displayed image?''. For texts, 
partial correctness was included in our accuracy metric. 
Although the correctness of the descriptions was subjective, 
most participants’ answers were binary: accurate/specific 
(e.g., “ice rock climbing” and “person in red hiking 
mountain”), or no comprehension. We
measured a shoulder surfer's or intended user's recognition  
rate (binary accuracy) as the percentage of text,
images, and videos correctly labeled by the participants. 
We also asked participants to indicate the percentage of 
text they could read. The participants were asked to 
complete a system usability scale (SUS) 
survey~\cite{brooke1996sus} regarding the quality of images 
and text from the intended user's perspective. 

% \footnote{All study materials will be anonymous and available at \url{https://osf.io/haxzb/?view_only=8629a44c79d74b71b603be3830f60583}}

% \input{FiguresTex/fig_userstudy}

\subsection{User Study Interview}
\label{subsec:method_interview}

Finally, we discussed various topics related to \name and 
received a variety of qualitative feedback. We obtained 
qualitative feedback (\cref{subsec:results_qualitative}) after 
the user study experiments and the usability questionnaires 
were completed. We kept the interview open-ended and casual to 
learn about participants' initial perceptions of the system and 
to note ideas or opinions we would otherwise have missed. We 
transcribed and summarized the participants' discussion points 
and aggregated them by counting the number of participants who 
discussed similar topics. To help guide and continue the 
discussion, we asked participants about several topics. The 
questions we asked can be found in \cref{table:qualitative}. \footnote{The exact phrasing when asking about these questions 
differed from participant to participant, but the overall ideas 
and topics remained consistent.}

% \newpage

\section{Evaluation Results} \label{sec:results}

\noindent We conducted the experiments and evaluations described in 
\cref{sec:methodology}, answering the following key questions. 
\begin{itemize}[leftmargin=0.4cm]
\item \textbf{Experiment 1:} \textit{How effective is our approach at 
protecting content such as colored images, videos, text, 
and mobile UIs?} In the experimentation with perceptual similarity metrics, 
\name acts as expected, by very closely ($SSIM > 0.9$) resembling the 
blurred/pixelated content at smaller sizes and the original content at 
the original size. Cloud-based OCR and image recognition models are 
only able to recognize 8.45\% of images and 3.16\% of texts protected by \name. 
Through user study tests, both in-person and crowdsourced, we find 
shoulder surfers are only able to recognize 25.00\% of images and 18.78\% of text. 
The crowdsourced study indicates that shoulder surfers can only recognize 32.24\% -- 35.50\% 
of the content within images, videos, and texts. 
Our experiments indicate also \name provides additional protection beyond using solely a privacy film. (\cref{subsec:results_similarity,subsec:results_semantic,subsec:results_mturk,subsec:results_similarity,subsec:privacy_film})

\item \textbf{Experiment 2:} \textit{Does \name meet real-time constraints 
for its screen hiding algorithms?} \name is capable of achieving smooth 
performance (43 FPS) at even the highest screen resolutions 
($3088 \times 1440$). At lower resolutions, \name can achieve 60+ FPS, 
the optimal performance for screens with a 60Hz refresh rate. While running, 
\name consumes an acceptable amount of memory and CPU while having a 
moderate impact on power consumption. (\cref{subsec:results_performance})

\item \textbf{Experiment 3:} \textit{Is the usability cost of applying \name 
to users' smartphones acceptable?} Users who are bothered and uncomfortable
with shoulder surfing are more likely to use \name, with users on average
reacting positively towards the quality of text and images. The usability of
the system was deemed above average, with a SUS score of 68.86. Our interviews with 15 participants provide additional insights into the usability of \name, discussing concerns related to eye-strain, 
activation, and use cases.
(\cref{subsec:results_mturk,subsec:results_usability,subsec:results_qualitative})
\end{itemize}

\subsection{Perceptual Similarity} \label{subsec:results_similarity}

\begin{figure}[t!]
\centering
\begin{subfigure}{\columnwidth}
  \centering
  \includegraphics[width=\columnwidth]{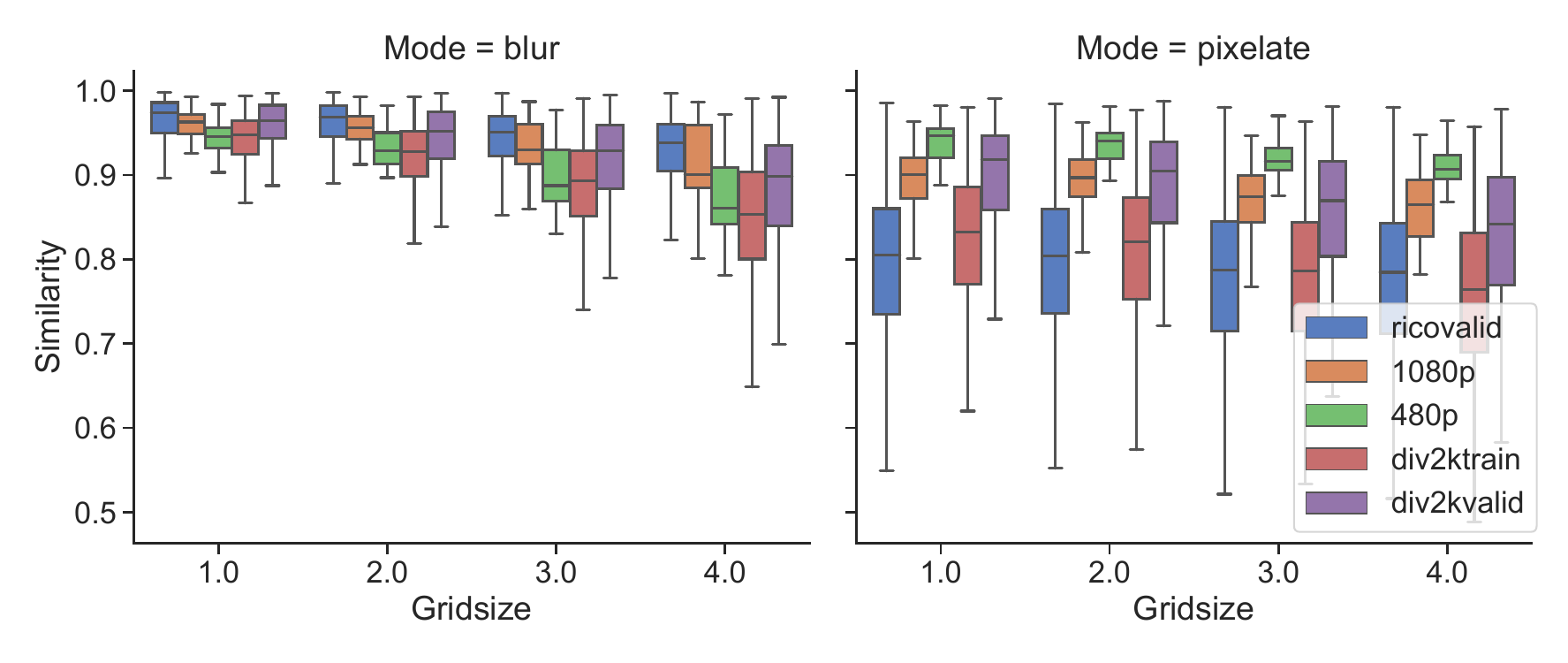}
  \caption{SSIM scores with different grid sizes.}
  \label{fig:ssim_gridsize}
\end{subfigure}
\begin{subfigure}{\columnwidth}
  \centering
  \includegraphics[width=\columnwidth]{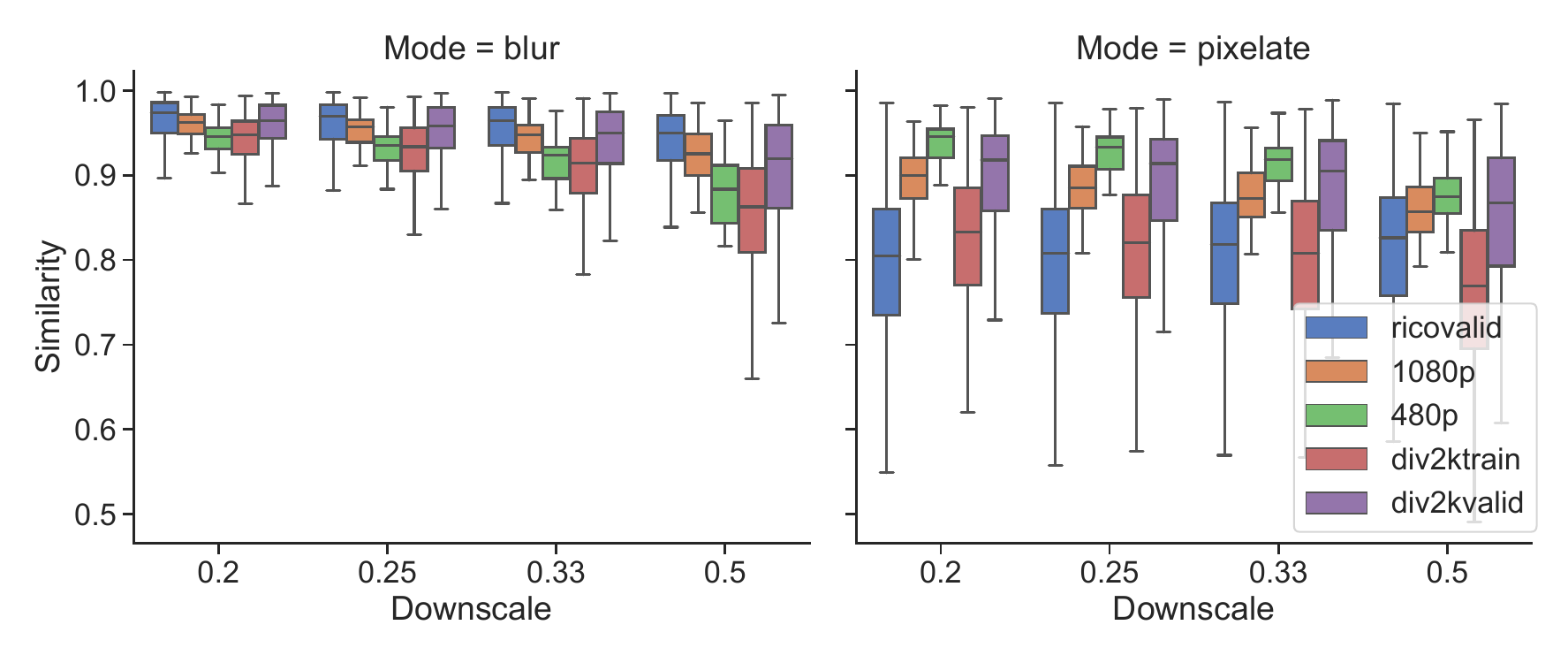}
  \caption{SSIM scores with different downscaling sizes.}
  \label{fig:ssim_downscale}
\end{subfigure}
\begin{subfigure}{\columnwidth}
  \centering
  \includegraphics[width=\columnwidth]{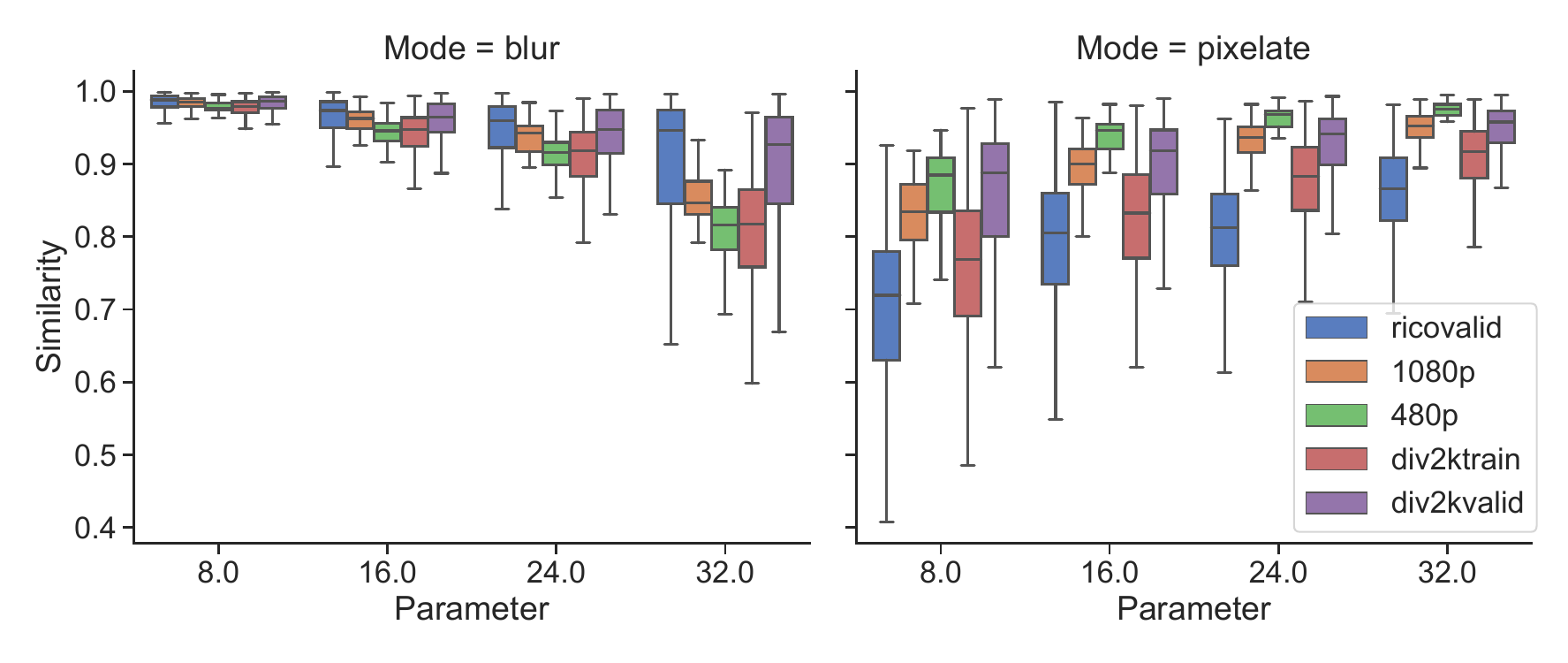}
  \caption{SSIM scores with different blurring intensities.}
  \label{fig:ssim_parameter}
\end{subfigure}
\caption{SSIM scores for each evaluated dataset while varying grid size, downscaling size, and blurring/pixelation intensity. SSIM scores are computed by comparing the downscaled protected image and blurred/pixelated image.}
\label{fig:ssim}
\end{figure}

\cref{fig:ssim} shows the measured SSIM scores, where scores near 1.0 indicate
high structural similarity, and scores near 0.0 indicate low similarity. 
We test for several parameters, such as grid size, 
downscaling size, and the window size for blurring. These $4 \times$ area 
downscaled images are consistent with the images presented to our MTurk 
participants in \cref{subsec:results_mturk}. Our algorithm was found to closely
mimic the blurred images with blurring intensity of up to $\sigma=24$, with the performance starting to degrade at $\sigma=32$ (\cref{fig:ssim_parameter}). SSIM scores are $> 0.9$, demonstrating the high efficacy of \name. Likewise, the structural similarity using pixelation degrades when the number of pixelated blocks is $< 16$. The results in 
\cref{fig:ssim_downscale} also suggest that \name can provide some 
protection even at distances of only 20'', though distances of 30'' and 
beyond afford the best protection guarantees. Finally, while the smallest 
grid size achieves the best performance, larger grid sizes are more 
effective at hiding larger text fonts.

\begin{figure*}[t!]
\centering
\begin{subfigure}{0.32\textwidth}
  \centering
  \includegraphics[width=\textwidth]{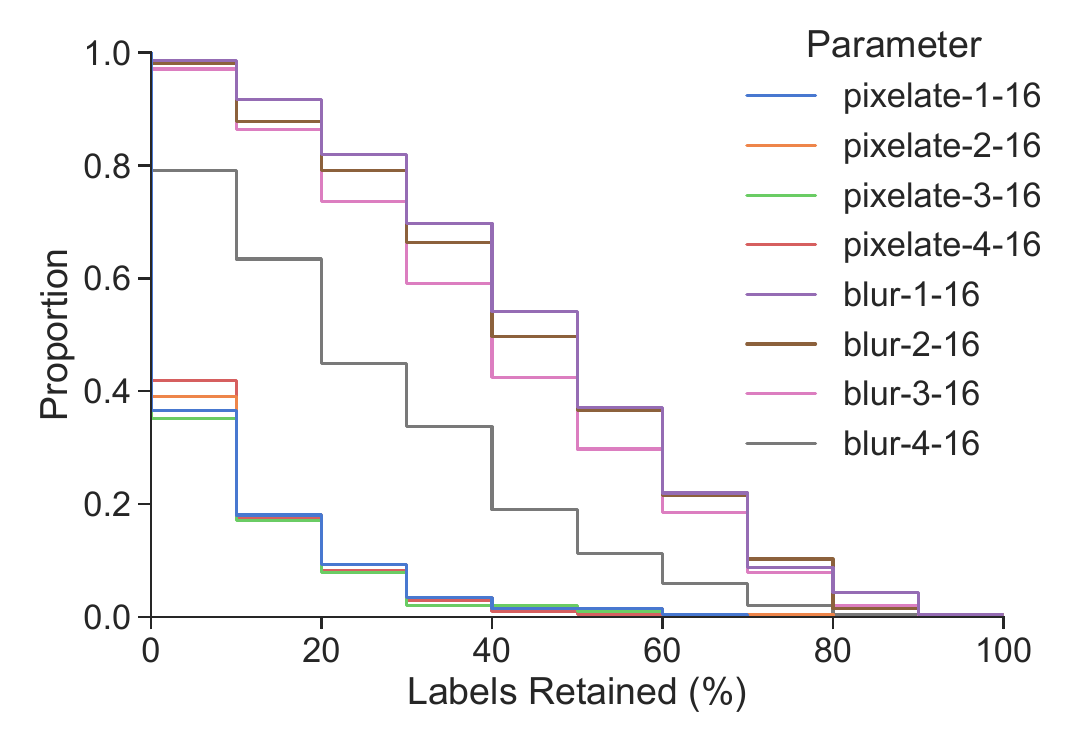}
  \caption{DIV2K Train Dataset}
  \label{fig:google_div2ktrain}
\end{subfigure}
\begin{subfigure}{0.32\textwidth}
  \centering
  \includegraphics[width=\textwidth]{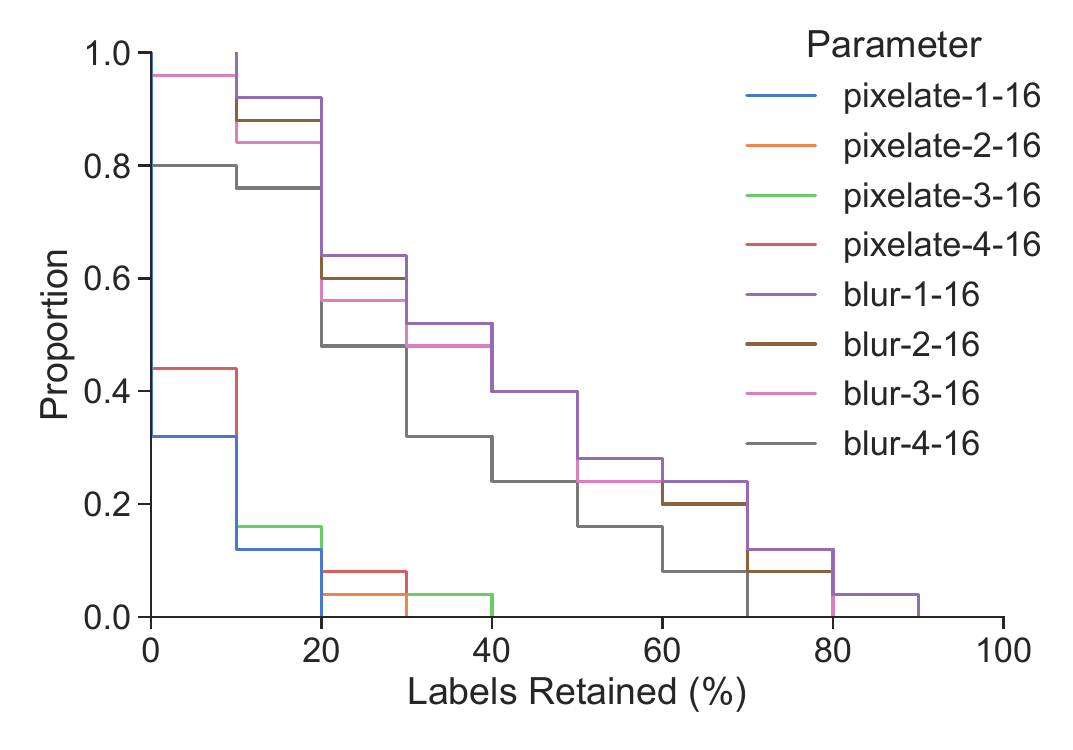}
  \caption{DIV2K Valid Dataset}
  \label{fig:google_div2kvalid}
\end{subfigure}
\begin{subfigure}{0.32\textwidth}
  \centering
  \includegraphics[width=\textwidth]{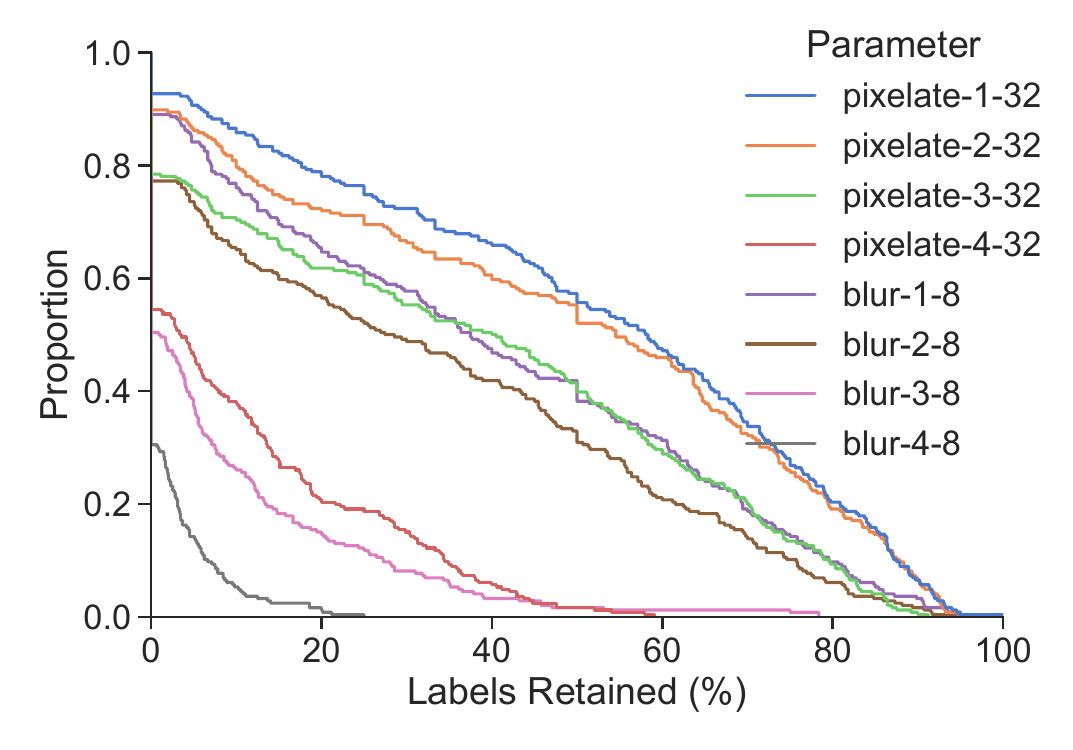}
  \caption{RICO Dataset}
  \label{fig:google_ricovalid}
\end{subfigure}
\caption{Results of using Google Cloud Vision API on protected images from 3 datasets. \cref{table:image_count} provides total image counts.}
\label{fig:google}
\end{figure*}

\subsection{Semantic Performance} \label{subsec:results_semantic}

\name is capable of reducing the detection rate of both image recognition 
and OCR systems. From an evaluation using the Google Cloud Vision API,
\cref{fig:google} shows how a majority of the protected images retain fewer 
than 50\% of the detected content compared to the original images. With
certain parameters, around 20\% of the evaluated images retain 0\% of the
originally detected labels. This observation holds for each 
evaluated dataset. In total, the DIV2K Valid, DIV2K Train, and RICO datasets 
protected by \name retain 22.93\%, 24.60\%, and 47.71\% of the original labels 
and text across all parameters. The high-resolution images protected using
pixelation preserve much more information than the images protected
using blurring. For images of mobile app UIs, blurring degrades the 
performance of the OCR system more than pixelation. When using blurring for 
text and pixelation for images, \name-protected images retain 8.45\%, 9.48\%, 
and 42.38\% for the DIV2K Valid, DIV2K Train, and RICO datasets. 
At larger grid sizes, the original text retained by the protected images decreases 
to as small as 3.16\%. These results indicate that \name may function best when 
switching between blurring and pixelation, depending on the type of interface 
presented on the device's screen. Selecting between blurring and pixelation can 
be made possible by detecting the amount of text on-screen using features such as 
OCR or using a phone's accessibility suite to determine the quantity of on-screen text. 
These results suggest that both shoulder surfers and vision systems may struggle 
with recognizing the information on a screen protected with \name.

\subsection{Performance Evaluation} \label{subsec:results_performance}

\begin{figure*}[t!]
\centering
\begin{subfigure}{0.24\textwidth}
  \centering
  \includegraphics[width=\textwidth]{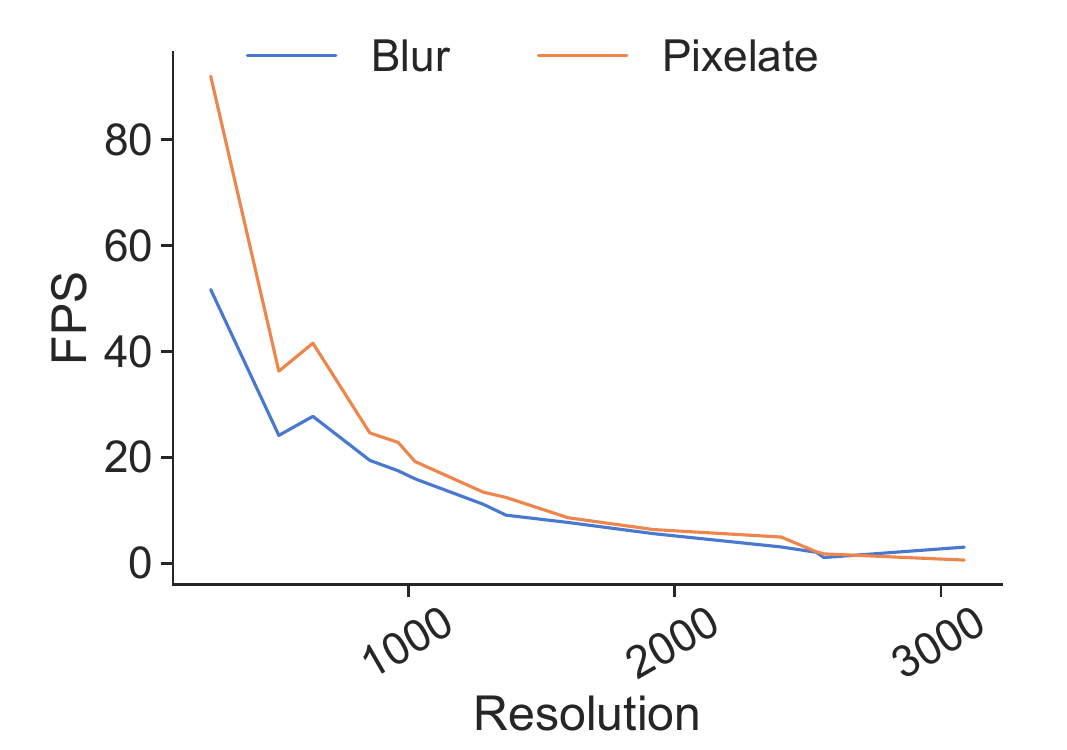}
  \caption{CPU-Only FPS}
  \label{fig:cpu_fps}
\end{subfigure}
\begin{subfigure}{0.24\textwidth}
  \centering
  \includegraphics[width=\textwidth]{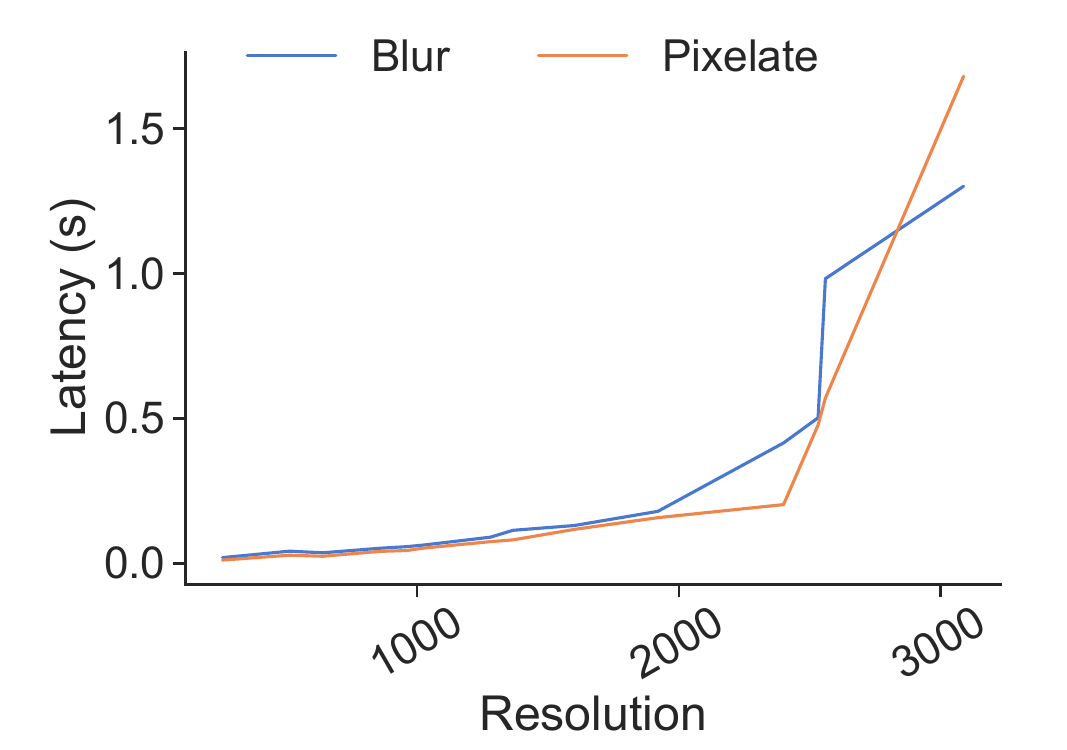}
  \caption{CPU-Only Latency}
  \label{fig:cpu_latency}
\end{subfigure}
\begin{subfigure}{0.24\textwidth}
  \centering
  \includegraphics[width=\textwidth]{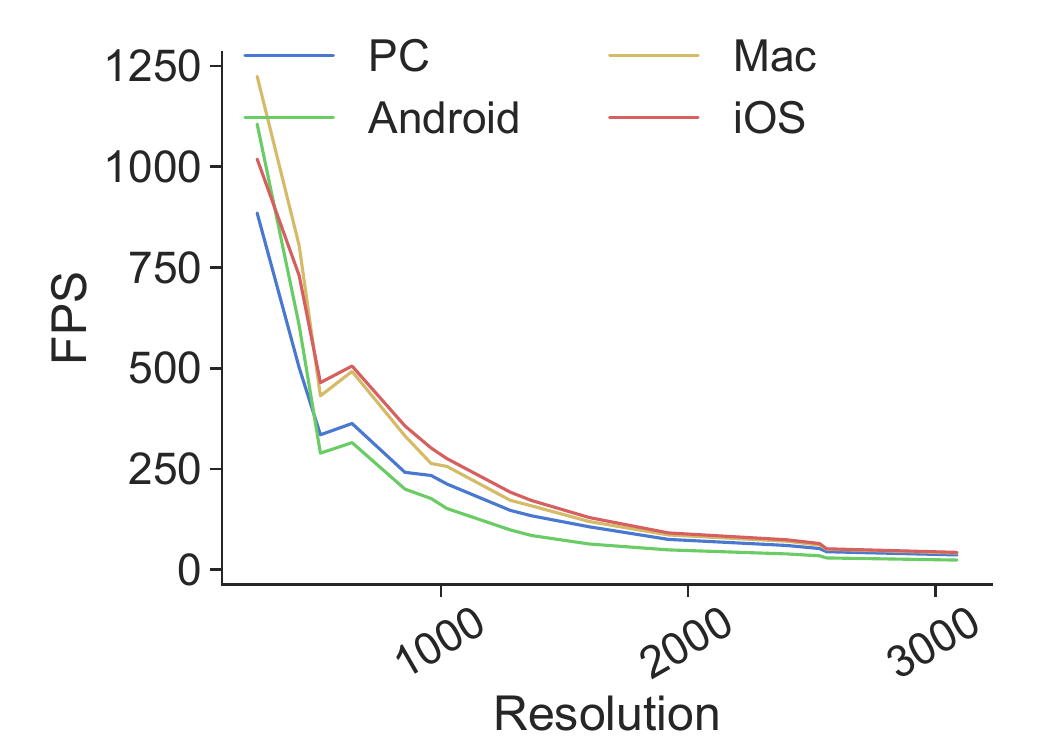}
  \caption{With-GPU FPS}
  \label{fig:gpu_fps}
\end{subfigure}
\begin{subfigure}{0.24\textwidth}
  \centering
  \includegraphics[width=\textwidth]{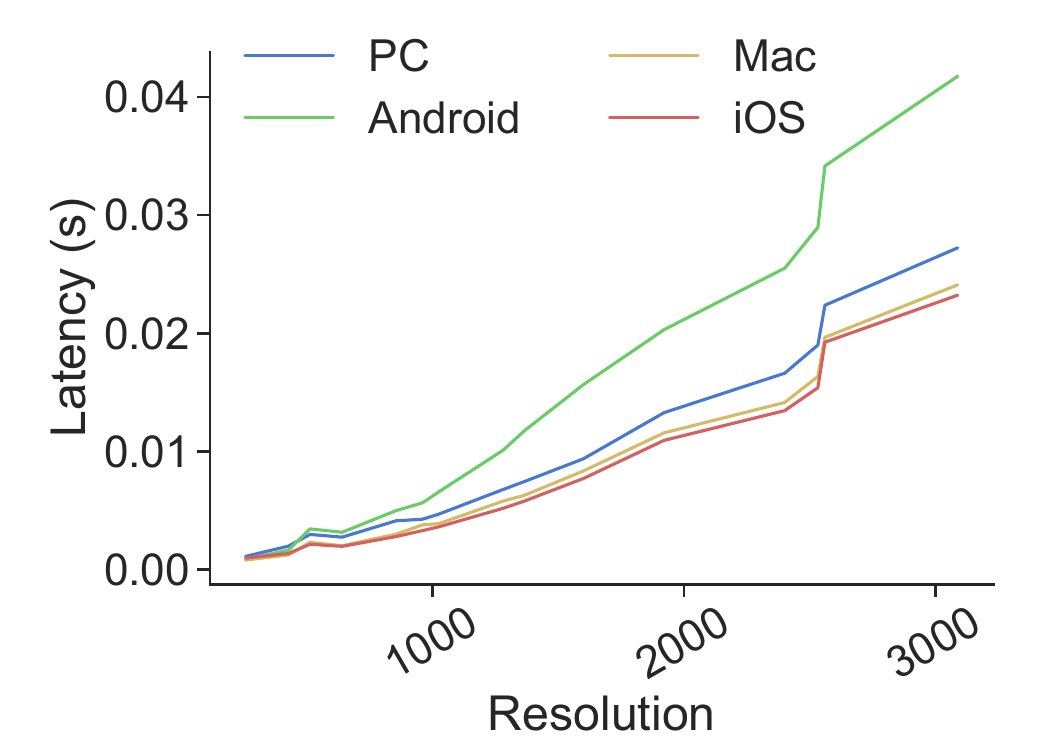}
  \caption{With-GPU Latency}
  \label{fig:gpu_latency}
\end{subfigure}
\caption{Measured FPS and latency using CPU-only and with-GPU implementations.}
\label{fig:performance}
\end{figure*}

As \name was designed to protect mobile device screens with real-time 
constraints in mind, we conducted a set of performance evaluations 
aimed at measuring \name's performance on a variety of screen sizes, 
video resolutions, and image sizes. 
% The processing time for \name using purely CPU achieves 0.51 FPS for screens of the largest resolution size ($3088 \times 1440$), 6.32 FPS for $1920 \times 1080$ videos, and 22.88 FPS for $854 \times 480$ videos. 
The processing time for \name with GPU support achieves 36.72 FPS, 75.27 FPS, 
and 241.72 FPS for resolution sizes of $3088 \times 1440$, $1920 \times 1080$, 
and $854 \times 480$ (\cref{fig:performance}).
At an average latency of 3.457ms 
on $512 \times 512$ \textit{full-color} images, the Android 
implementation of \name provides a $487 \times$ speed-up compared to 
the 1684ms achieved on \hidescreen with $512 \times 512$ 
\textit{grayscale} images. The iOS, MacOS, and PC implementations of 
\name achieve 2.153ms, 2.317ms, and 2.986ms, respectively. The average 
latency of running \name on an image or frame of $3088 \times 1440$ is 
27.23ms (PC), 41.75ms (Android), 24.10ms (MacOS), and 23.23ms (iOS). 
% Even the CPU-only implementations of \name have an average latency of 
% 45.79ms ($37 \times$ speed-up) on $512 \times 512$ colored images. 
As shown in \cref{fig:performance}, \name can achieve an average 
of 43 FPS on $3088 \times 1440$ screen resolutions using the iOS 
implementation. This performance would allow the screen protected by 
\name to update according to the user's inputs at an effective refresh rate 
of 43Hz. Considering most smartphones run at a 60Hz refresh rate, 
our implementation provides an acceptable trade-off for increased 
privacy. Alternatively, users who prefer responsiveness to 
high resolution may display the protected screen at a lower 
resolution to attain $>60$ FPS. Our evaluation shows \name 
achieves a $>60$ FPS at screen resolutions of $1170 
\times 2532$ and smaller on iOS (\cref{table:mobile_fps}).
The trade-offs between latency and
screen resolution are similarly present in energy consumption and
memory usage. In the case where most users prefer stable performance
over higher screen resolution, \name will prioritize selecting
a screen resolution that can match the device's refresh rate.

\begin{figure*}[t!]
\centering
\begin{subfigure}{0.33\textwidth}
  \centering
  \includegraphics[width=\textwidth]{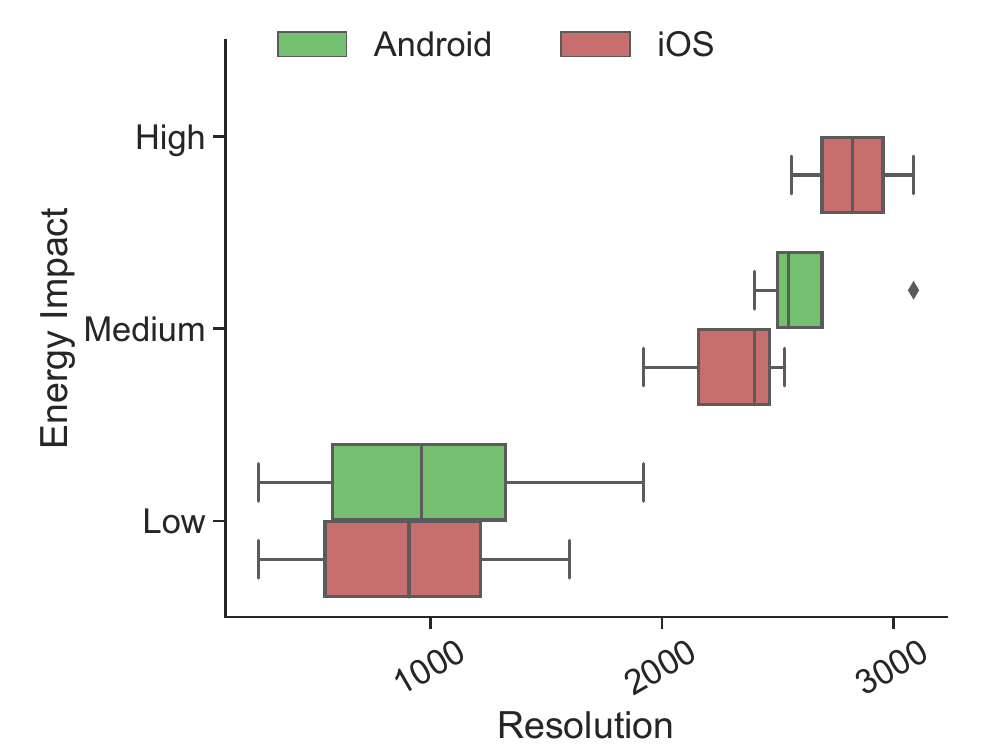}
  \caption{Energy Impact}
  \label{fig:energy}
\end{subfigure}
\begin{subfigure}{0.33\textwidth}
  \centering
  \includegraphics[width=\textwidth]{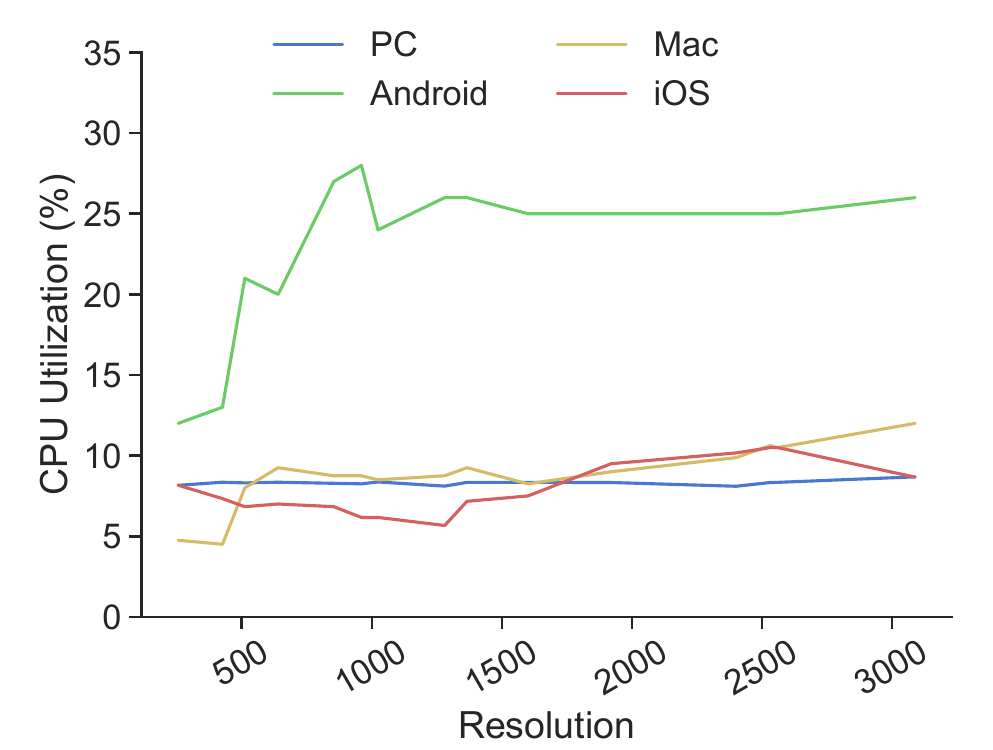}
  \caption{CPU Utilization}
  \label{fig:utilization}
\end{subfigure}
% \begin{subfigure}{0.24\textwidth}
%   \centering
%   \includegraphics[width=\textwidth]{Figures/cpu_memory.pdf}
%   \caption{CPU-Only Memory \\(Desktop)}
%   \label{fig:cpu_memory}
% \end{subfigure}
\begin{subfigure}{0.33\textwidth}
  \centering
  \includegraphics[width=\textwidth]{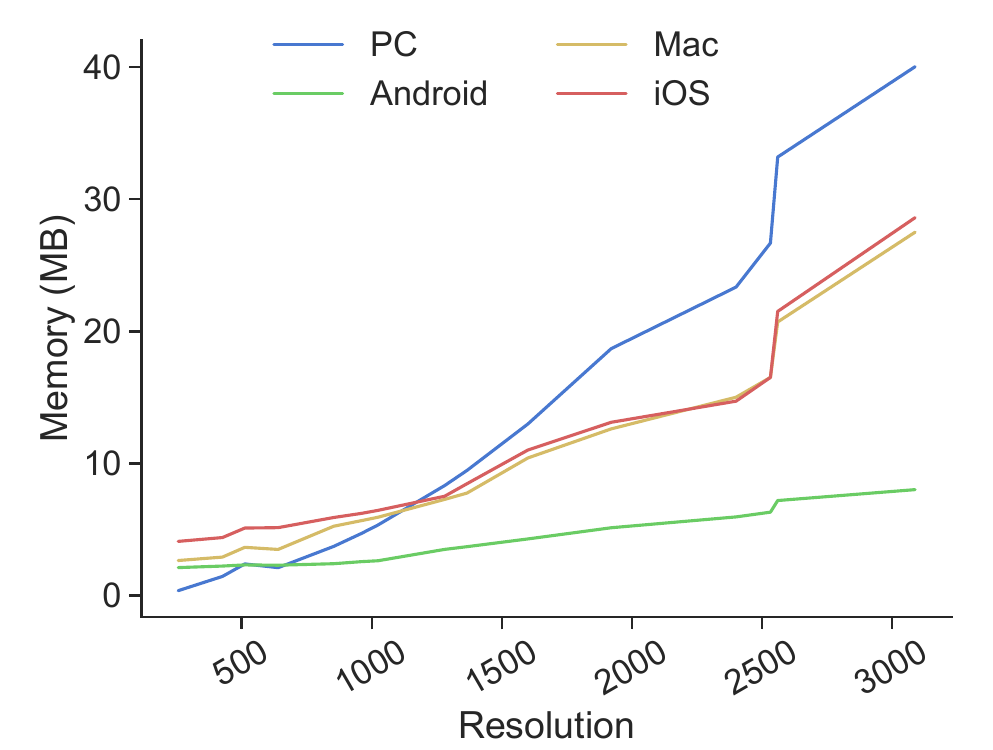}
  \caption{With-GPU Memory}
  \label{fig:gpu_memory}
\end{subfigure}
\caption{Resource consumption such as energy usage, CPU utilization, and memory usage are measured and reported.}
\label{fig:resources}
\end{figure*}

Since \name must be lightweight enough to run on mobile devices,
we ensure that the memory usage is small enough to not cause significant
memory errors or stuttering. 
% Figs.~\ref{fig:resources}d and c
\Cref{fig:gpu_memory}
% and \cref{fig:cpu_memory} 
shows the memory usage of with-GPU and non-GPU 
implementations of \name across each platform.
The memory usage scales with the resolution size, reaching up to 
40.04MB per frame in the GPU implementation of \name. 
Using only the CPU consumes more memory with each frame (up to 299.95MB).

On mobile devices, \name is 
observed to achieve low energy 
consumption for resolution sizes of up to $1920 \times 1080$. 
For larger resolution screen sizes, the measured energy impact 
is medium to high (see \cref{fig:energy}). 
The maximum measured CPU utilization of \name for each of 
the four evaluated devices is shown in \cref{fig:utilization}. 
Utilization is a percentage of all cores on the device (\cref{table:devices}).
For the most part, utilization of all the CPU cores on 
the device remains below 25\%. 
Overall, especially at lower resolutions, the measured resource 
usage and estimated energy impacts indicate that \name is 
lightweight enough to be run on mobile devices without 
over-taxing battery life or causing 
overheating/thermal issues. 

\subsection{MTurk Study} 
\label{subsec:results_mturk}

The main purpose of our MTurk study is a large-scale 
evaluation of \name's efficacy in defending against shoulder 
surfing. Our study demonstrates that shoulder surfers can only 
recognize as low as 32.24\% of the text on our protected images from 
an effective distance of 41'' (\cref{table:mturk_accuracy}). 
As a comparison, shoulder surfers can recognize up to 83.84\% of the 
unprotected text from the same distance. 
This degrades the shoulder surfers' recognition rate by 51.60 percentage points. 
Our results for images and videos achieve similar protection 
improvements, with decreases in recognition rate of 60.75 
and 61.61 percentage points, respectively. 
These reductions in recognition rate demonstrate \name's 
potential for reducing the amount of information a shoulder 
surfer can glean from an unwitting user.

\begin{table}[!t]
\caption{MTurk and in-person content recognition rates.}
\begin{center}
\resizebox{0.95\columnwidth}{!}{
  \begin{tabular}{c c c c c}
    \toprule
    {Platform} & {\bf Setting} & {\bf Text(\%)} & {\bf Image(\%)} & {\bf Video(\%)} \\ 
    \midrule
    MTurk & Intended User (Protected) & 97.22 & 81.75 & 79.80 \\
     & Shoulder Surfer (Original) & 83.84 & 96.25 & 96.46 \\
     & Shoulder Surfer (Protected) & 32.24 & 35.50 & 34.85 \\
    \midrule
    In-Person & Intended User (Protected) & 98.05 & 82.58 & 86.64 \\
     & Shoulder Surfer (Original) & 77.27 & 100.0 & 100.0 \\
     & Shoulder Surfer (Protected) & 15.91 & 24.24 & 47.04 \\
     & Shoulder Surfer (Original, $45 \degree$) & 100.0 & 100.0 & 100.0 \\
     & Shoulder Surfer (Protected, $45 \degree$) & 21.90 & 22.22 & 51.67 \\
    \bottomrule
    \end{tabular}
}
\end{center}
\label{table:mturk_accuracy}
\end{table}

% We also measured the participants' response times in recognizing 
% and answering each survey question about the protected and 
% unprotected images. For images on average, there is a 4.23-second difference, or a 37.77\% increase in answering time. 
% On average for text, responses are 6.30 seconds slower, 
% implying a 46.93\% increase in answering time. Participants 
% took longer to respond for protected images and text than 
% the unprotected ones. \Cref{fig:mturk_timings} depicts the
% distribution of participants' response times for different 
% content types.

% \begin{figure}[t]
%   \centering
%   \includegraphics[width=\columnwidth]{Figures/mturk_qualities.pdf}
%   \caption{Likelihood of using \name conditioned on responses from \cref{fig:mturk_privacy}.}
%   \label{fig:mturk_qualities}
% \end{figure}

\begin{figure*}[t]
% \begin{subfigure}{0.58\columnwidth}
%   \centering
%   \includegraphics[width=\columnwidth]{Figures/mturk_timings.pdf}
%   \caption{Response Times}
%   \label{fig:mturk_timings}
% \end{subfigure}
\begin{subfigure}{0.7\textwidth}
  \includegraphics[width=\textwidth]{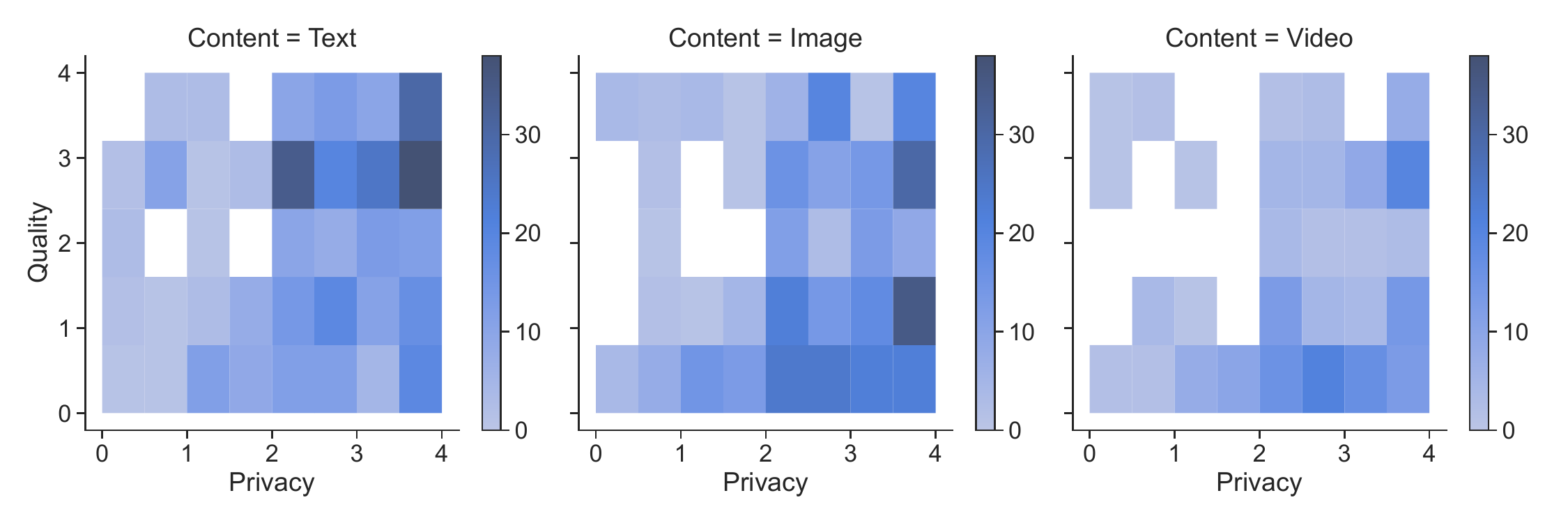}
  \caption{Likelihood of using \name conditioned on responses from \cref{fig:mturk_privacy}.}
  \label{fig:mturk_qualities}
\end{subfigure}
\begin{subfigure}{0.23\textwidth}
  \centering
  \includegraphics[width=\textwidth]{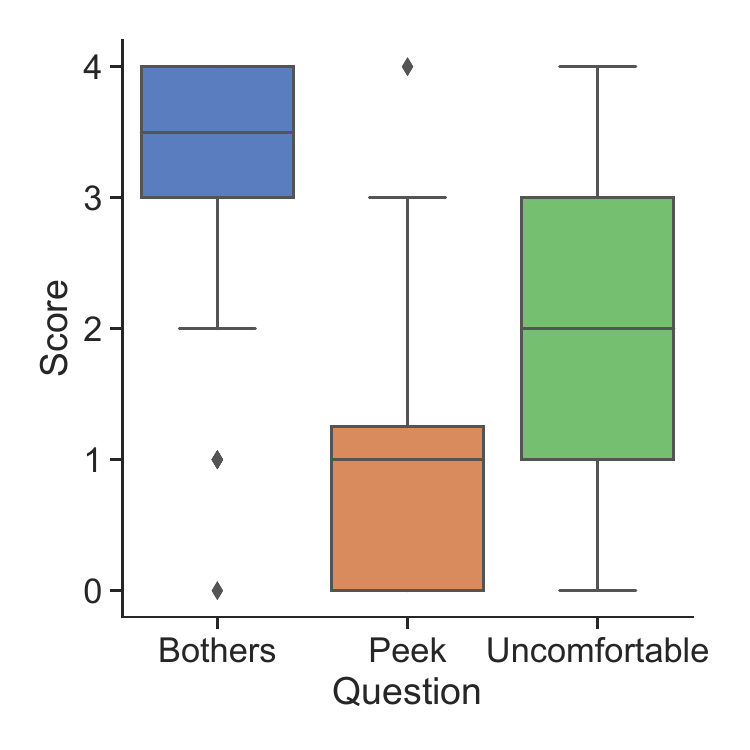}
  \caption{Privacy Attitudes}
  \label{fig:mturk_privacy}
\end{subfigure}
\caption{(a) Likelihood of using \name conditioned on responses from (b). (b) Distribution of MTurk privacy responses.}
\label{fig:mturk_other}
\vspace*{-0.2in}
\end{figure*}

% \begin{figure}[t]
%   \centering
%   \includegraphics[width=0.70\columnwidth]{Figures/mturk_privacy.pdf}
%   \caption{Participants' attitudes towards shoulder surfing. 5-point Likert responses to 1) how bothered users are by shoulder surfing, 2) frequency of peeking at others' screens, and 3) how uncomfortable users are with using their smartphone in public settings.}
%   \label{fig:mturk_privacy}
% \end{figure}

We also assessed our participants' perceptions of shoulder 
surfing. The mean 5-point Likert scores for 1) how bothered 
users were by others peeking at their phones, 2) how often users 
peeked at others' phone screens in public, and 3) how uncomfortable 
users were with looking at their phones in public
areas were 3.28, 1.02, and 1.83, respectively. 
Fig.~\ref{fig:mturk_privacy} shows that the participants were 
generally averse to being involved in shoulder surfing
both as the victim and adversary.

We gathered participants' feedback on their contentedness of the quality of images using 
\name for different content types conditioned on their attitudes 
towards shoulder surfing. For those who were both 
bothered by shoulder surfing and uncomfortable with using 
their smartphones in public 
settings, the mean Likert score for the likelihood of using the 
screen protection in public settings was 2.61, 2.21,
and 2.00 for mobile UIs, images, and videos, respectively. 
These indicate that, on average, privacy-conscious 
participants were mostly happy with using \name in public 
settings to protect their privacy. Fig.~\ref{fig:mturk_qualities} depicts 
how, especially for images of mobile app UIs protected using blurring, participants 
were happy with the quality of images. The density plots indicate a correlation 
between concern with shoulder surfing and the likelihood of using \name.
Participants were less satisfied with the image quality for the 
high-resolution images and videos protected using pixelation.

\subsection{User Study} \label{subsec:results_usability}
We assessed the recognition rate of screens protected by \name 
for an in-person setting and observed an overall recognition rate
of 26.94\% for shoulder surfers. For users close to the screen as
the intended user, the recognition rate is 89.57\%. For text
visibility, we observed stronger protection with a shoulder
surfer recognition rate of around 5.88\%. 
Close to the screen, almost 100\% of the text is
visible to the intended user. 
\Cref{table:mturk_accuracy} shows how in-person
participants were only able to recognize 15.91\% of the protected
texts and 24.24\% of the protected images. The video domain represented 
a much more challenging problem, as participants were still able
to recognize the scenes 47.04\% of the time
As a comparison, without the protection provided by \name, participants
could clearly see and recognize almost 80\% of the texts and 100\% of 
the images and videos.

\begin{table}[!t]
\caption{Overall in-person recognition rates}
\vspace*{-0.1in}
\begin{center}
\resizebox{0.95\columnwidth}{!}{
  \begin{tabular}{c c c}
    \toprule
    {\bf Setting} & {\bf Recognition Rate (\%)} & {\bf Text Visible (\%)} \\ 
    \midrule
    Intended User (Protected) & 89.57 & 94.05 \\
    Shoulder Surfer (Original) & 90.37 & 62.96 \\
    Shoulder Surfer (Protected) & 26.94 & 5.876 \\
    Shoulder Surfer (Original, $45 \degree$) & 100.0 & 96.82 \\
    Shoulder Surfer (Protected, $45 \degree$) & 27.84 & 14.67 \\
    \bottomrule
    \end{tabular}
}
\end{center}
\label{table:inperson_accuracy}
\vspace*{-0.1in}
\end{table}

\begin{figure}[h]%[t]
  \centering
  \includegraphics[width=\columnwidth]{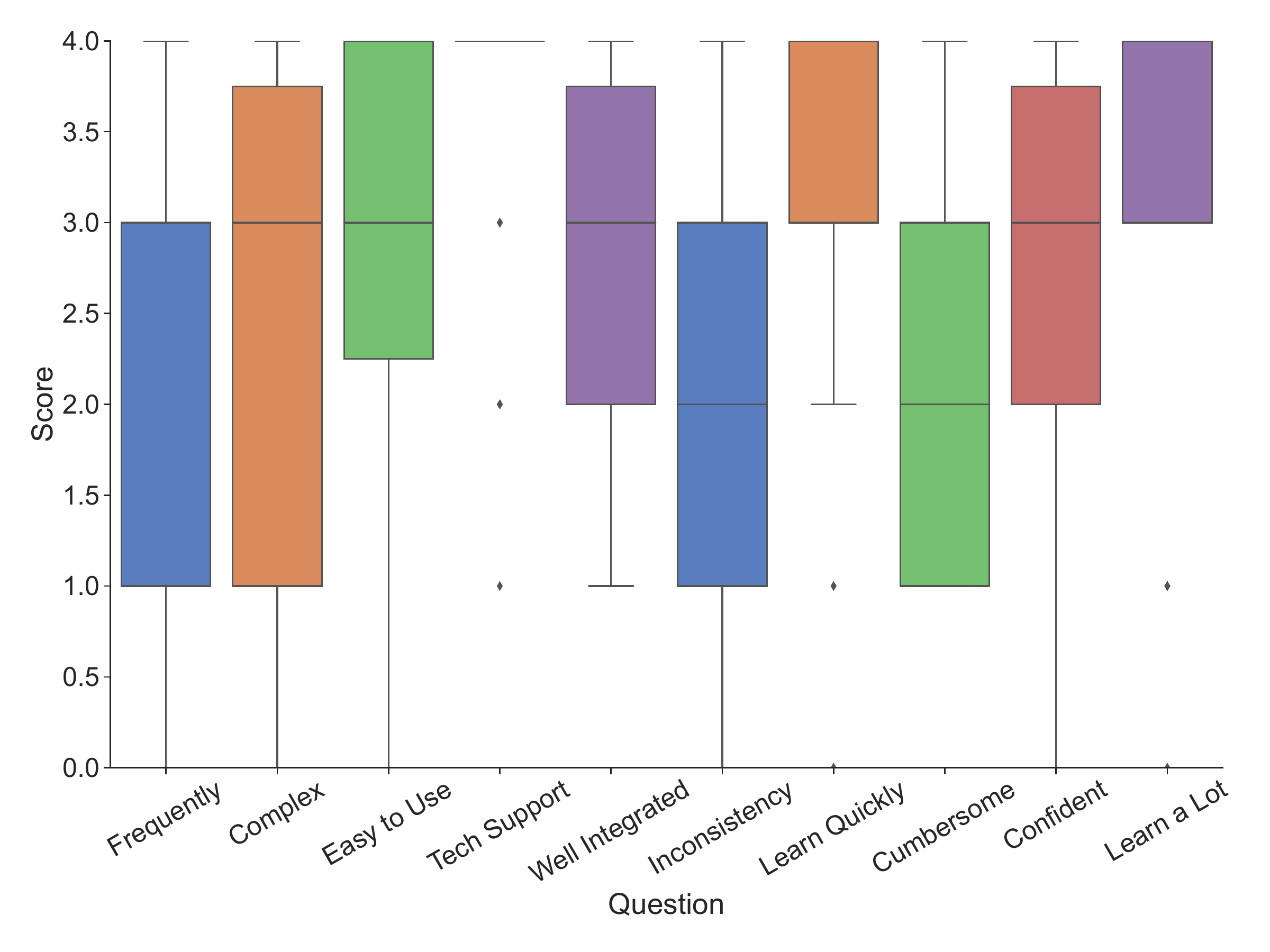}
  \caption{Normalized SUS scores for each question.}
  \label{fig:inperson_sus}
\end{figure}

After answering questions about the content displayed on \name, 
our participants responded to a SUS questionnaire. The average SUS 
score of \name was 68.86, where a SUS score above 68 is deemed
above average~\cite{brooke1996sus}. The distribution of responses is
presented in \cref{fig:inperson_sus}. 
These SUS scores include the original 7 participants along with the 15 new participants after the study design changed. Participants responded negatively
mainly to the consistency and cumbersomeness of \name, due to certain types 
of content being more easily recognizable than others.

While we did not explicitly measure the time taken for 
participants to comprehend the information displayed, 
we observed that all participants took longer to respond 
when viewing the protected screen than the unprotected screen 
from the same distance of 41'' and 20'' with a 45 $\degree$ angle.

% We received several points of criticism as well. Some participants stated that ``It was really difficult to see some of the pixelated images'', ``It would be more difficult to use with my trifocals or someone with gradient glasses'', and ``If it was something I needed to read very quickly, I would be a bit annoyed.''}
% Another participant observed that while it was
% difficult to recognize some of the pixelated images and videos
% even at close distances, it would help if they were the owners 
% of the device and using it with context.

% Participants also identified additional use cases where 
% Participants bothered by shoulder surfing stated that they would take privacy-preserving actions: e.g., using something to cover their phone, lowering brightness, using face recognition or fingerprint authentication, being aware of their surroundings, using a privacy film, or just stop using their phone entirely.

\subsection{Qualitative Feedback}
\label{subsec:results_qualitative}

Overall, out of 15 participants, 8 participants indicated 
they would be uncomfortable with shoulder surfers peeking 
at their devices (Q7), 7 participants indicated they preferred 
using \name over using a privacy film (Q2), 7 participants found 
an option to set the blurring intensity to be useful (Q11), and 
6 participants stated that the ability to toggle \name 
would be helpful. (Q11) 
7 participants stated they would use \name for protecting 
financial data and PIN entry (Q3), and 3 participants said 
they would use it to protect personal texts and photos (Q4). 
7 participants found the blurring that \name introduces 
to the user to be slightly annoying (Q5), and only 3 
participants stated they had eyestrain as the intended user (Q6). 
Overall, these results support our claim that \name would be 
helpful for protecting privacy-conscious users who are concerned about shoulder surfing (more in \cref{table:qualitative}).

Generally, participants wanted \name for PIN entry, and some 
participants wanted \name to activate automatically for 
certain apps with more sensitive information (Q11). Participants 
also indicated that zooming, leaning in, and increasing 
brightness improved the usability of \name as the intended 
user (Q13). 
We also observed generational differences in our responses, for 
example, older participants generally used their phone less 
than younger participants in public and found less of a need 
for \name. One participant suggested that \name may perform 
differently for other languages (Q14).

% Participants also identified additional use cases where \name would be useful in preventing shoulder surfing: e.g., their child trying to figure out the PIN to play games, using their phone in church or lecture halls, using their phone at tourist spots or unfamiliar locations, having to use the phone for time-sensitive tasks in public, commuting in larger cities, and having the device font size larger. 

Some participants were excited and wanted to see \name implemented on their smartphones, while others did not see themselves ever personally using \name. Overall, participants reacted positively towards \name, noting that it was very difficult to see the protected on-screen information from the perspective of a shoulder surfer.

\subsection{Comparison with Privacy Films}
\label{subsec:privacy_film}

We purchased several privacy films for use in our evaluations 
and user studies.
% \footnote{\url{https://www.amazon.com/dp/B08RP9G74R}, \url{https://www.amazon.com/dp/B09BZ8QLWZ}} 
During our in-person user study, we also conducted several experiments comparing the angle of readable text between the privacy film and \name. Participants could not read any text on the screen protected by the privacy film from the 45$\degree$ angle, although they could easily see everything whenever the phone was tilted sideways. When we asked participants to lean over towards the screen, they only needed to lean over an average of 4.55". After activating \name with the privacy film still on, we asked participants to lean over, and they needed to lean an average of 10.55". Thus, combining \name and privacy films provides the best protection. We also observed that the efficacy of privacy films was dependent on the environmental brightness relative to the screen brightness. 
For example, they are not effective at darkening the screen in dark settings, nor at 100\% screen brightness levels, especially on laptops. Additionally, privacy films generally cost \$10--\$30 and are designed for specific device screen sizes/types. 
Several participants indicated annoyance at the inability of disabling the physical privacy film for smartphones. 
For example, one participant noted that the screen would appear dark and blurry if the phone was turned slightly, 
although they would consider using the film for large stationary devices like laptops. 
We performed an additional evaluation with the Google OCR service and observed that privacy films 
provide no protection at angles of $\leq n30 \degree$, whereas \name still provides some protection. 
We note that privacy films provide protection against optical zoom whereas \name is not. 
\Cref{table:film} depicts these results, demonstrating that in some settings, \name protects 
more information than a privacy film, and using both \name and a privacy film offers 
complete protection in the evaluated settings.

% \newpage

\section{Discussion} 
\label{sec:discussion}

\subsection{Implications of Findings}

Our experimental evaluation indicates the feasibility of 
implementing a software-based privacy film for 
mobile device screens. Having a widely accessible 
low-latency screen protection mechanism
could increase users' awareness of shoulder 
surfing attacks and preserve their
privacy without significantly disrupting their 
device usage. Without the need for additional 
physical components, \name could be 
implemented agnostic of both apps and devices. Privacy-conscious users would no longer need to purchase and install new films whenever they change mobile devices, averaging once every 22.7 months for American adults (and more frequently for
younger users)~\cite{smartphoneupgrades2022}.
Highly cautious users will find increased privacy 
guarantees by applying both \name and a privacy film
to protect their on-screen information. 

\subsection{Prototype User Interface}

As shown in \cref{fig:prototype}, we have developed a low-fidelity 
prototype for toggling \name on both the iOS and Android 
operating systems. \name would be most naturally implemented 
as a toggle-able widget, with more advanced users being able to 
adjust individual parameters and features in the device's 
system settings and preferences. Users can manually toggle \name upon entering public settings. We expect most users to activate \name before viewing private or sensitive content. For most users, the default 
parameters can be set to $gridsize=1$ and blurring with 
$\sigma=8$, which achieves the best overall performance in 
the evaluation performed in \cref{subsec:results_semantic}. 
The default screen resolution size can be set to the maximum 
size in which the system achieves 60 FPS. Some users may opt 
to adjust these parameters to attain higher screen resolution 
or increased protection guarantees. We consider several pre-designed parameter settings:

\begin{enumerate}[noitemsep,leftmargin=0.4cm,topsep=5pt]
    \item High protection [low-brightness, low-contrast, high-blur]
    \item Med protection [med-brightness, med-contrast, med-blur]
    \item low protection [high-brightness, high-contrast, low-blur]
\end{enumerate}

\subsection{Limitations} 
 \label{subsec:limitations}

%\name suffers from several limitations. 
The most notable limitation of \name 
is the loss of some information and image quality. 
For example, users responded more negatively to pixelation
than blurring, despite pixelation providing stronger
privacy guarantees. Likewise, the blurring intensity and
pixelation block size can affect interpretability
and readability.
Nonetheless, \name still allows users to continue with 
their intended applications with minimal interruptions 
to their workflow. This trade-off between usability and 
privacy guarantees is inevitable when simultaneously providing 
intended users with their information and protecting 
information from shoulder surfers in the same 
screen.
Additionally, 
shoulder surfers will still be able to infer certain 
high-level details. Several participants in 
our in-person user study were able to guess the type of
mobile app being displayed and the general category of images. However,
they were unable to interpret or read any specific details. 
% In our MTurk study, the recognition
% rate for shoulder surfers was low, but not zero.
% This is a trade-off inherent to our design, as it prioritizes
% low latency, near 100\% recognition rates for the
% intended users, and universal protection across
% all types of content.

% \name cannot provide adequate protection to high-contrast pairs of colors. This imperfect protection 
% against shoulder surfers can leak some information. 
% One potential remedy for high-contrast content can be 
% to restrict the complementary pair of colors within 
% the range of viable colors (see \cref{fig:colors}). 
% This would incur a small amount of overhead in the form 
% of adding a few extra matrix operations to the 
% existing \name's algorithm (\cref{algorithm}).

% Additionally, the brightness of the device screen can 
% have an impact on \name's efficacy. We observed that 
% moderate and low levels of brightness provide additional
% protection for screens using \name.

% \name affords begin to falter. \name is most effective 
% when used with moderate and low brightness settings.

People of older age or those with visual impairments, vision degradation, or color-blindness may also find it slightly more challenging to use \name. Our user study included 6 participants with minor visual impairments who were able to read and recognize almost all displayed content as the intended user. For users with severe visual impairments, mobile OSs provide accessibility APIs for toggling color filters and increasing contrast. Through preliminary tests, it is possible to use \name with color-blindness filters with little to no impact on the comprehension of content. However, increasing contrast will negatively impact the protection guarantees. Regarding the risk of eyestrain, the length of the in-person study totaled about 60 minutes, with time spent staring at the screen averaging around 30 minutes. We offered the participants at several instances a chance to take a break, but none of the participants required a break. Only 3 participants noted experiencing minor eyestrain as the intended user. However, it is unclear whether using \name for prolonged periods of time ($>$1 hour) will cause eyestrain for users due to looking at blurry/pixelated content.

% It may be possible to address some of these problems 
% at the cost of protection guarantees by reducing 
% \name's blurring or pixelation intensity or adding a 
% color-blindness filter before applying \name.

\subsection{Future Work} \label{subsec:future_work}

We have presented and demonstrated a 
shoulder surfing protection mechanism, and left 
integration of \name into Android OS and iOS as 
ongoing/future work. Creating an adaptive 
solution to adjust parameters such as blurring and pixelation intensity
and grid size 
is another technical challenge which 
requires further investigation. 

Other directions for future work include assessing the 
efficacy of \name using various device screen sizes
and other blurring or pixelation methods. Finding the 
optimal trade-off between usability and privacy in each 
of \name's parameters is another useful topic to explore.

% \newpage

\section{Conclusion} \label{sec:conclusion}

We have presented and evaluated \name to prevent the shoulder 
surfing of information displayed on mobile devices.
It is designed to protect all types of on-screen information 
--- text, colored images, mobile app UIs, videos, and 
smartphone browsing --- {\em in real time}, without significantly 
hampering the user's interactions with the mobile device.
\name can be regarded as a software version of a privacy film. 
By blurring/pixelating a screen, generating a checkered grid, 
and computing complementary colors, \name can generate images that 
appear readable and interpretable at close distances, 
but appear blurry and pixelated at distances of 30'' and angles of 45'' and beyond.
Having a software-based defense against shoulder surfing built 
into devices can increase user awareness of shoulder surfing and prevent 
adversaries from accessing/stealing sensitive information. 
\name was designed and implemented as a low-cost and easily-adoptable 
defense mechanism against shoulder surfing.
%\name achieved our goals of protecting on-screen information from shoulder surfers, 
%operating in real time, and being minimally intrusive to the intended user. 
%For example, 
% \name was able to reduce recognition rates to 25.00\% and 18.78\% 
% for image and text content in our in-person study, and reduces recognition rates to 35.50\% and 32.24\%in our MTurk study. 
% \name also protects up to 96.84\% of information from being 
% properly detected and recognized on the Google Cloud Vision API. 
% \name was implemented on several platforms and achieved high 
% frame rates for maximum-sized screen resolutions (Android = 24 FPS,
% iOS = 43 FPS). Users willing to use smaller screen resolutions 
% can use \name with lower latency (Android = 49 FPS, iOS = 91 FPS).
% \name also operates with acceptable memory usage, CPU utilization, 
% and energy overhead, consuming moderate energy at high resolutions. 
% Finally, our user studies demonstrated that \name protects screens 
% at an acceptable cost for privacy-conscious users. Those who were bothered 
% or uncomfortable with shoulder surfing found the text and image quality 
% acceptable for use in public settings. The in-person participants 
% found the system to be easily usable, with an average SUS score of 79.57.

\section*{Acknowledgments}
The work reported in this paper was supported in part by the Army Research Office (ARO)
under Grant No.~W911NF-21-1-0057. Special thanks to the shepherd, the RTCL members, and all study participants for their incredibly helpful feedback!
\section*{Availability}
The code, datasets, and user study samples are provided at \url{https://github.com/byron123t/eye-shield}, and \url{https://www.bjaytang.com/projects/post_008/}.

% \balance
% \printbibliography
\bibliographystyle{IEEEtran}
\bibliography{references}
\appendix
\section*{Appendix}

\section{Contrast, Brightness, and Environment} \label{subsec:contrast_brightness}

\begin{figure*}[t!]
\centering
\begin{subfigure}{0.24\textwidth}
  \centering
  \includegraphics[width=\textwidth]{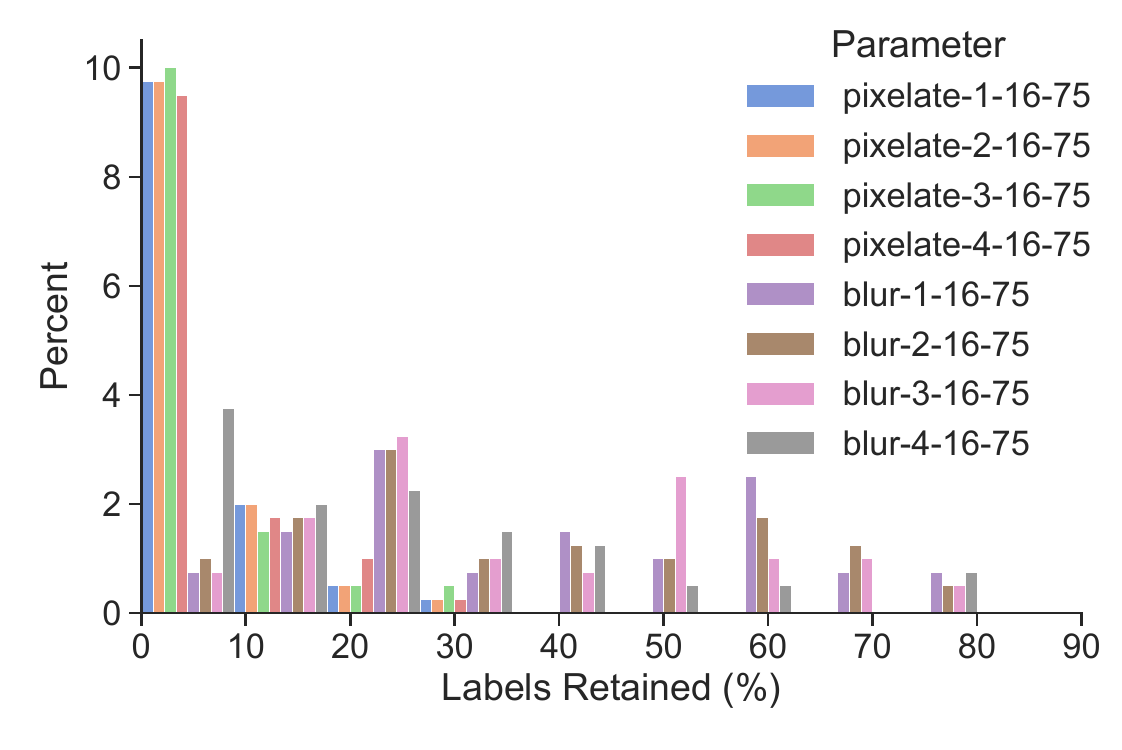}
  \caption{DIV2K, Low Contrast}
  \label{fig:google_contrast_div2kvalid_low}
\end{subfigure}
\begin{subfigure}{0.24\textwidth}
  \centering
  \includegraphics[width=\textwidth]{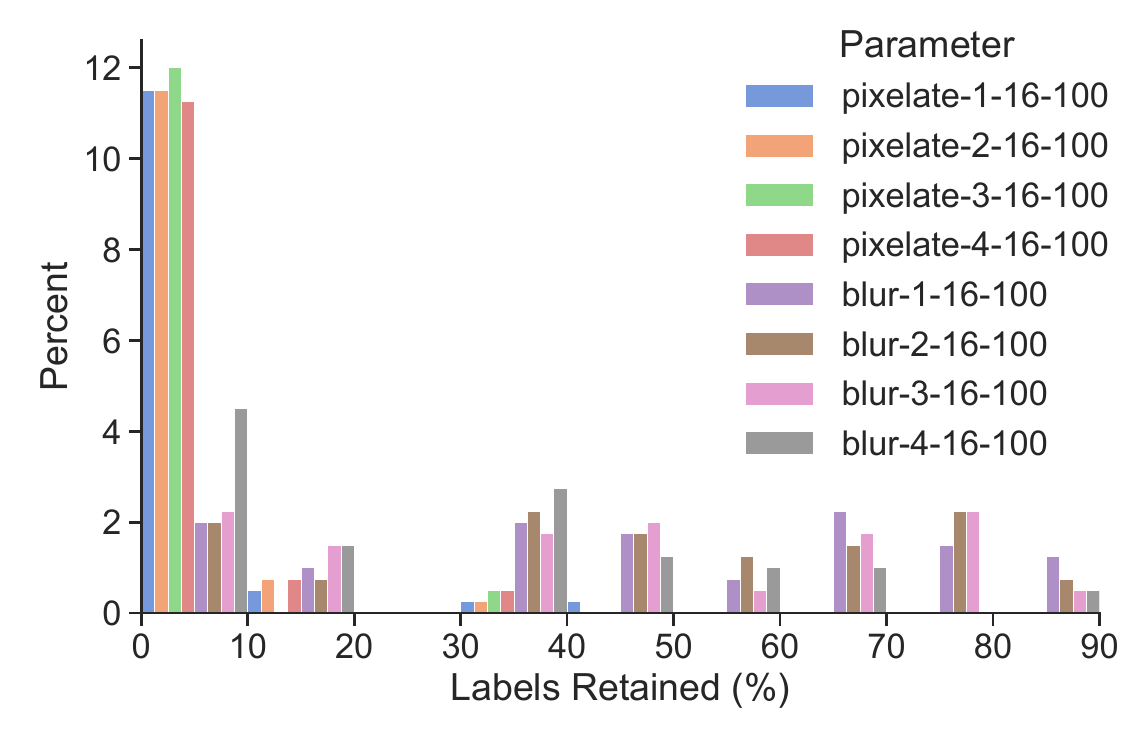}
  \caption{DIV2K, Moderate Contrast}
  \label{fig:google_contrast_div2kvalid_mod}
\end{subfigure}
\begin{subfigure}{0.24\textwidth}
  \centering
  \includegraphics[width=\textwidth]{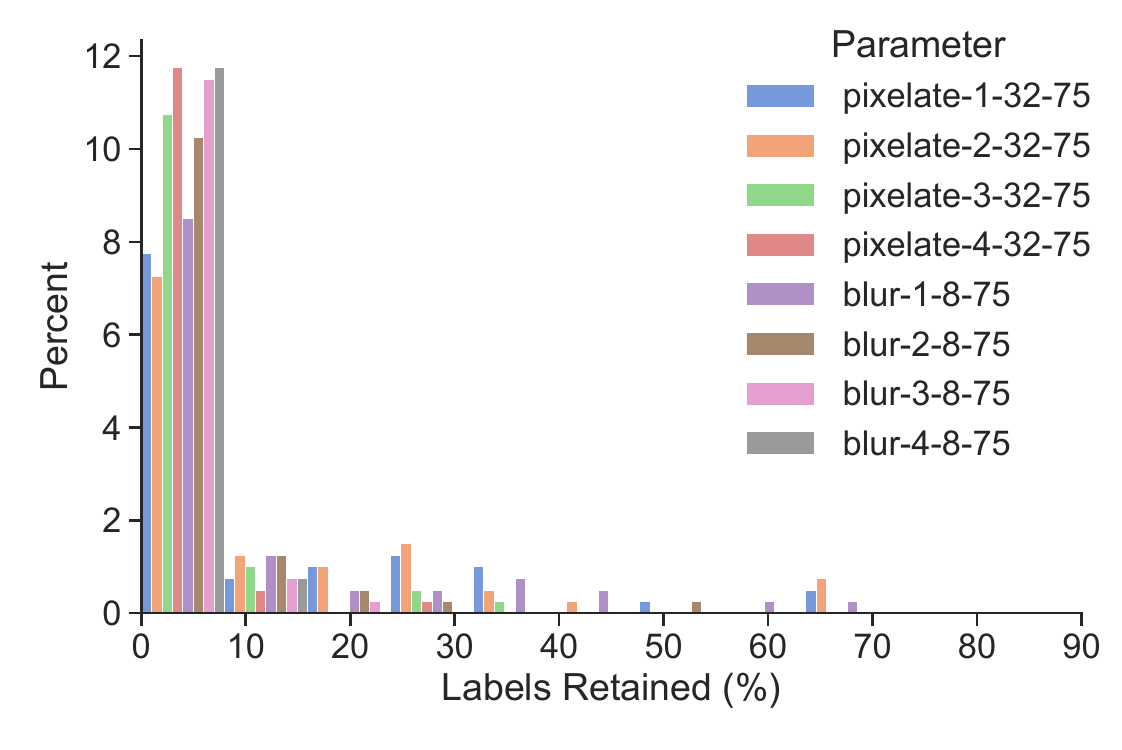}
  \caption{RICO, Low Contrast}
  \label{fig:google_contrast_ricovalid_low}
\end{subfigure}
\begin{subfigure}{0.24\textwidth}
  \centering
  \includegraphics[width=\textwidth]{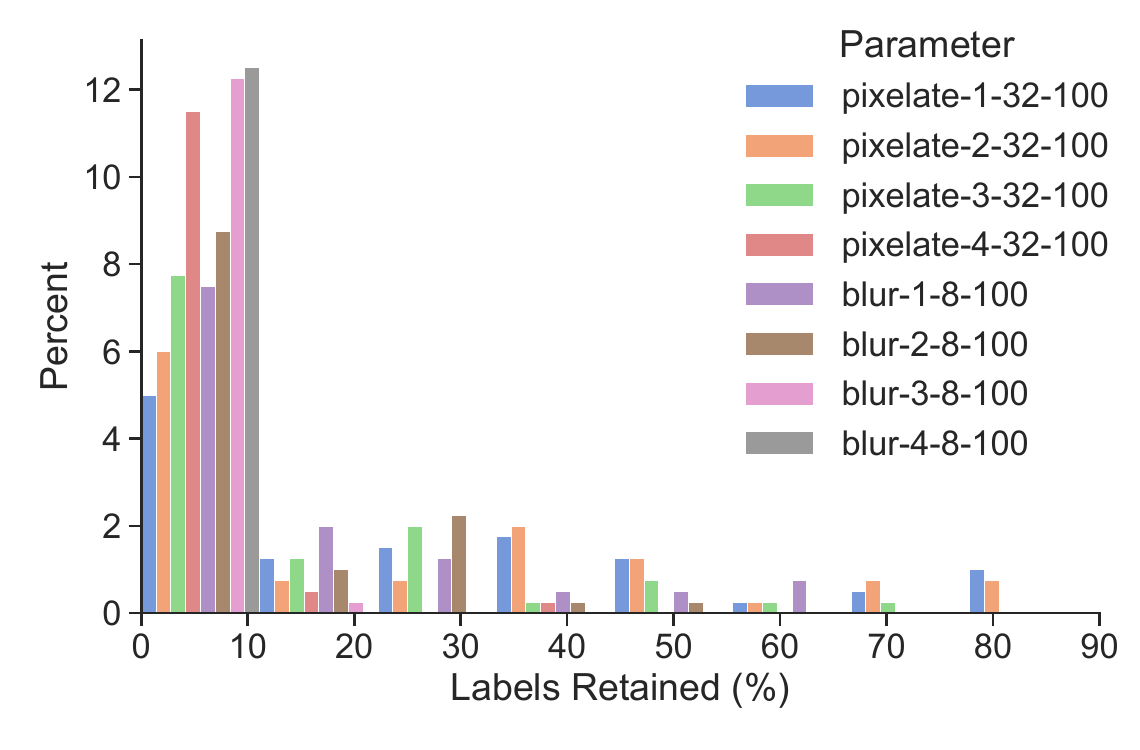}
  \caption{RICO, Moderate Contrast}
  \label{fig:google_contrast_ricovalid_mod}
\end{subfigure}
\caption{Results of using Google Cloud Vision API on protected images from 2 datasets with low contrast and moderate contrast settings. The unique images evaluated on each dataset total 50 images.}
\label{fig:google_contrast}
\end{figure*}

\begin{figure*}[t!]
\centering
\begin{subfigure}{0.098\textwidth}
  \centering
  \includegraphics[width=\textwidth]{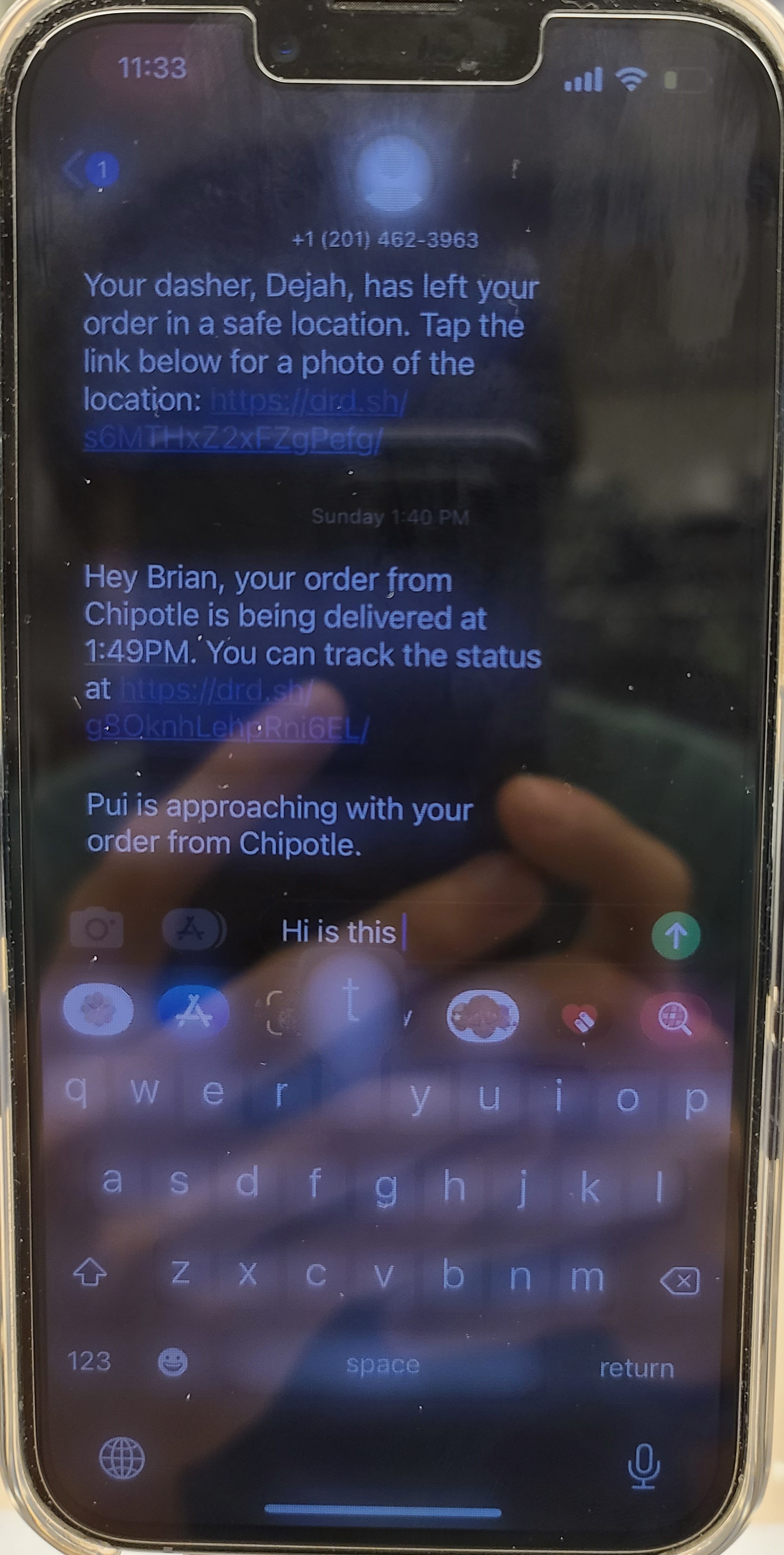}
  \caption{33\%}
  \label{fig:darkest_screen_close}
\end{subfigure}
\begin{subfigure}{0.098\textwidth}
  \centering
  \includegraphics[width=\textwidth]{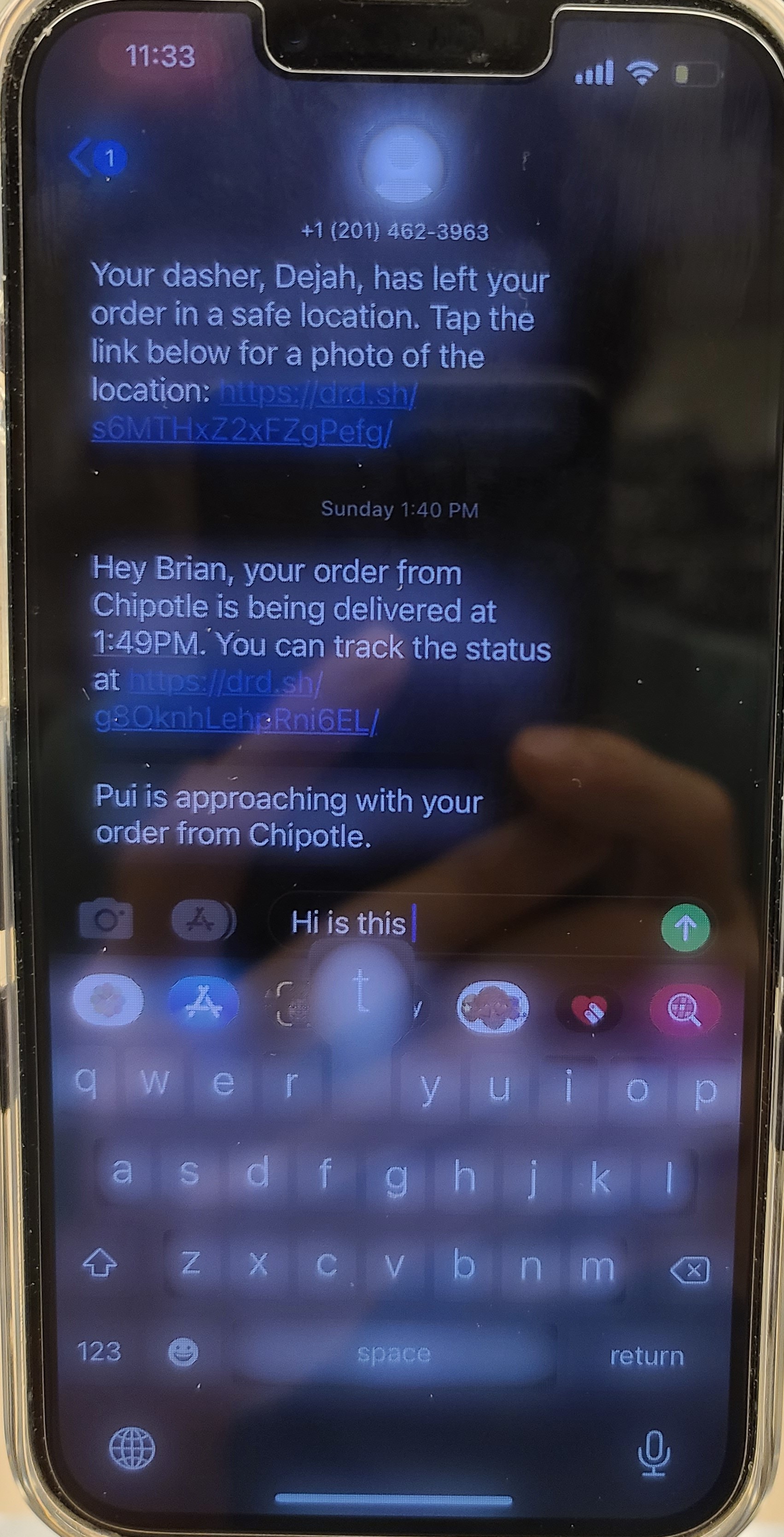}
  \caption{66\%}
  \label{fig:moderate_screen_close}
\end{subfigure}
\begin{subfigure}{0.098\textwidth}
  \centering
  \includegraphics[width=\textwidth]{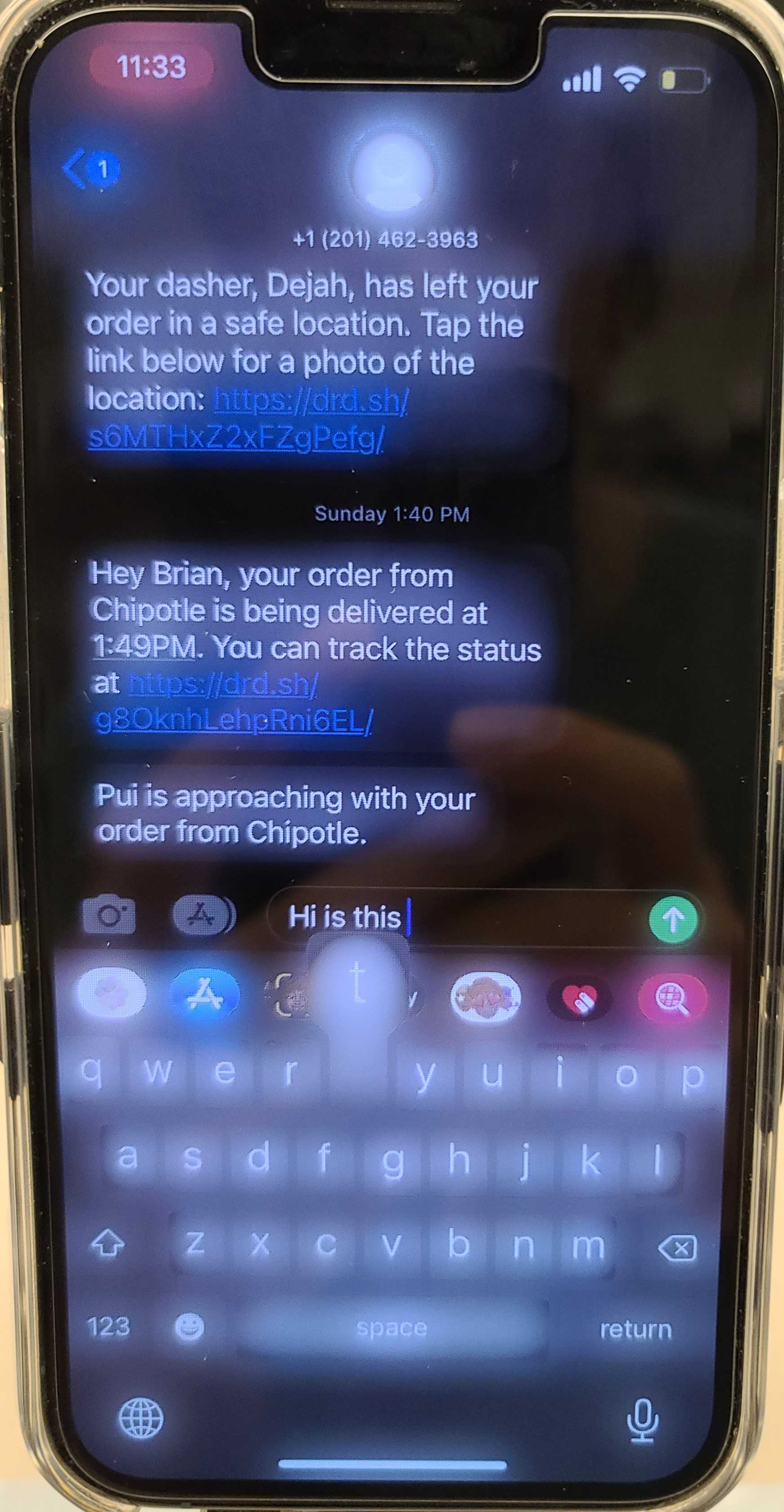}
  \caption{100\%}
  \label{fig:brightest_screen_close}
\end{subfigure}
\begin{subfigure}{0.188\textwidth}
  \centering
  \includegraphics[width=\textwidth,angle=270]{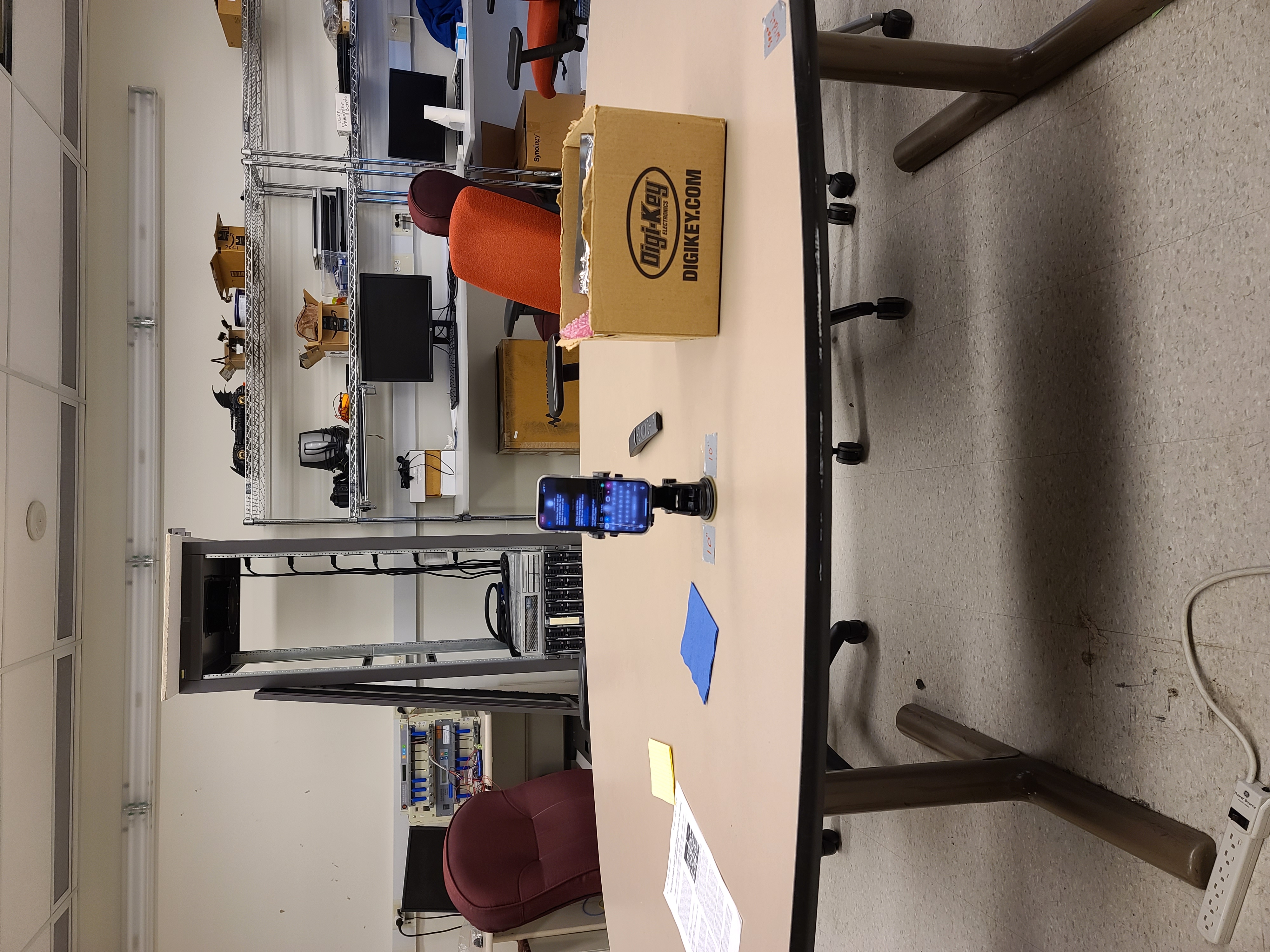}
  \caption{Bright Env}
  \label{fig:bright_lab}
\end{subfigure}
\begin{subfigure}{0.188\textwidth}
  \centering
  \includegraphics[width=\textwidth,angle=270]{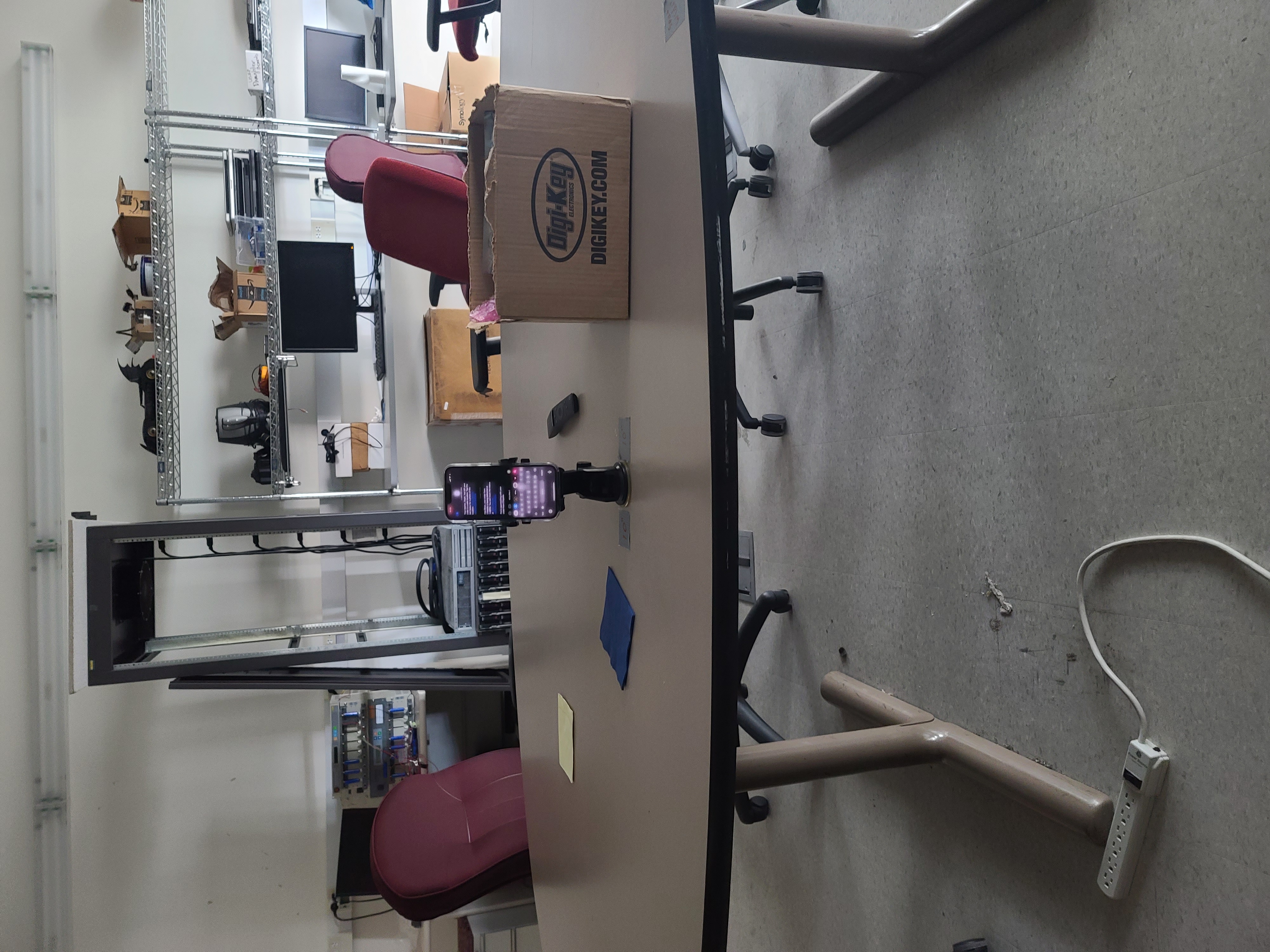}
  \caption{Dark Env}
  \label{fig:dark_lab}
\end{subfigure}
\caption{Device screen brightness from the intended user's perspective (a,b,c). Environment lighting experiment setup (d,e).}
\label{fig:brightness_environment}
\end{figure*}

We conducted an additional set of experiments to evaluate the efficacy of protecting on-screen information with high-contrast colors and different device and environmental brightness settings. We evaluated images protected using \name and displayed on a smartphone with 3 different device brightness settings (33\%, 66\%, and 100\%) in a lab with the lights on and off. \Cref{fig:brightness_environment} shows the experiment setup and the example photos used for the experiment. Note that at a close distance, there is little degradation in readability. Photos were taken of the device screen in all the described settings, and the cropped photos were evaluated on Google's OCR system to determine the impact of brightness on the efficacy of \name. The word detection rate for images taken from the intended user's perspective is only decreased by 14 words and 9.5 words in the dark and bright environments, respectively. From the shoulder surfer's perspective, the average word detection rate decreases by as much as 21 words from the side angle, and 8.25 words from the direct angle. While the intended user can still read most words at any brightness setting, in the darkest brightness settings, a shoulder surfer may be unable to read \textit{any} words on the protected screen (\cref{table:brightness}). Finally, we added a feature to decrease the contrast of the content displayed using \name and evaluated the impact of decreasing contrast for both OCR and image recognition. The impact of the new feature on performance was negligible. Upon generating protected images with low and moderate-low contrast, we observed the impact of the decreased contrast on the Google Cloud APIs. For texts, the percentage of protected content is greatly increased for all parameters, while images receive little benefit (\cref{fig:google_contrast}). This suggests that decreasing brightness and contrast is an effective method to further improve the protection rate for large fonts and high-contrast UIs.

\begin{table}[!t]
\caption{Results of using Google Cloud OCR on photos taken of a text message UI with several different protection mechanisms at 50\% screen brightness with $5 \times$ optical zoom.}
\begin{center}
\resizebox{\columnwidth}{!}{
  \begin{tabular}{c c c c}
    \toprule
    {\bf Environment} & {\bf Angle} & {\bf Protection} & {\bf Word Recognition Rate} \\ 
    \midrule
    Dark & 30 $\degree$ & \name & 87 \\
    Dark & 30 $\degree$ & Film & 83 \\
    Dark & 30 $\degree$ & Both & 0 \\
    Dark & 30 $\degree$ & None & 90 \\
    \midrule
    Dark & 45 $\degree$ & \name & 54 \\
    Dark & 45 $\degree$ & Film & 0 \\
    Dark & 45 $\degree$ & Both & 0 \\
    Dark & 45 $\degree$ & None & 80 \\
    \midrule
    \midrule
    Light & 30 $\degree$ & \name & 60 \\
    Light & 30 $\degree$ & Film & 74 \\
    Light & 30 $\degree$ & Both & 0 \\
    Light & 30 $\degree$ & None & 88 \\
    \midrule
    Light & 45 $\degree$ & \name & 48 \\
    Light & 45 $\degree$ & Film & 0 \\
    Light & 45 $\degree$ & Both & 0 \\
    Light & 45 $\degree$ & None & 70 \\
    \bottomrule
    \end{tabular}
}
\end{center}
\label{table:film}
\end{table}

\begin{table}[!t]
\caption{Results of using Google Cloud OCR on photos taken of a text message UI with several different brightness settings.}
\begin{center}
\resizebox{\columnwidth}{!}{
  \begin{tabular}{c c c c}
    \toprule
    {\bf Environment} & {\bf Angle} & {\bf Screen Brightness} & {\bf Word Recognition Rate} \\ 
    \midrule
    Dark & Close & Darkest & 42.50 \\
    Dark & Close & Moderate & 56.50 \\
    \midrule
    Dark & Far & Darkest & 0.500 \\
    Dark & Far & Moderate & 15.25 \\
    \midrule
    Dark & Side & Darkest & 0.000 \\
    Dark & Side & Moderate & 5.250 \\
    \midrule
    \midrule
    Light & Close & Darkest & 62.00 \\
    Light & Close & Moderate & 62.50 \\
    Light & Close & Brightest & 71.50 \\
    \midrule
    Light & Far & Darkest & 20.00 \\
    Light & Far & Moderate & 19.00 \\
    Light & Far & Brightest & 28.25 \\
    \midrule
    Light & Side & Darkest & 0.000 \\
    Light & Side & Moderate & 2.750 \\
    Light & Side & Brightest & 21.00 \\
    \bottomrule
    \end{tabular}
}
\end{center}
\label{table:brightness}
\end{table}

\section{Supplemental Experiment Setup Information}

In this section, we provide several additional experiment details. We describe the development stack of \name for each platform-specific implementation in \cref{table:stack}. Our stack consists of image processing and general purpose GPU processing libraries.

\begin{table}[h]
\caption{The development stack of for each platforms.}
\begin{center}
\resizebox{0.9\columnwidth}{!}{
  \begin{tabular}{c c c c}
    \toprule
    {\bf Platform} & {\bf Language} & {\bf Image Processing} & {\bf GPGPU} \\ 
    \midrule
    Windows PC & Python & OpenCV & CUDA 11.6 (CuPy) \\
    MacOS Laptop & Swift & CoreImage & Metal \\
    Android Smartphone & C++ & OpenCV & Vulkan (Kompute)\\
    iOS Smartphone & Swift & CoreImage & Metal \\
    \bottomrule
    \end{tabular}
}
\end{center}
\label{table:stack}
\end{table}

\begin{table}[t]
\caption{Mobile performance of \name on large screens.}
\begin{center}
\scalebox{0.8}{
  \begin{tabular}{c c c}
    \toprule
    {\bf Resolution} & {\bf Android (FPS)} & {\bf iOS (FPS)} \\ 
    \midrule
    $1920 \times 1080$ & 49.25  & 91.39  \\
    $1080 \times 2400$ & 39.20  & 74.29  \\
    $1170 \times 2532$ & 34.52  & 64.95  \\
    $2560 \times 1440$ & 29.27  & 51.95  \\
    $1440 \times 3088$ & 23.95 & 43.05 \\
    \bottomrule
    \end{tabular}
}
\end{center}
\label{table:mobile_fps}
\end{table}

\begin{table}[t]
\caption{A list of the image resolutions used in our evaluation.}
\begin{center}
\scalebox{0.65}{
  \begin{tabular}{c c c}
    \toprule
    {\bf Resolution} & {\bf Aspect Ratio} & {\bf Purpose} \\ 
    \midrule
    $256 \times 144$ & 16:9 & Video Resolution  \\
    $426 \times 240$ & 16:9 & Video Resolution  \\
    $640 \times 360$ & 16:9 & Video Resolution  \\
    $854 \times 480$ & 16:9 & Video Resolution  \\
    $960 \times 540$ & 16:9 & Video Resolution  \\
    $1024 \times 576$ & 16:9 & Video Resolution  \\
    $1280 \times 720$ & 16:9 & Video Resolution  \\
    $1366 \times 768$ & 16:9 & Video Resolution  \\
    $1600 \times 900$ & 16:9 & Video Resolution  \\
    $1920 \times 1080$ & 16:9 & Video Resolution  \\
    $2560 \times 1440$ & 16:9 & Video Resolution \\
    $512 \times 512$ & 1:1 & \hidescreen Comparison \\
    $1080 \times 2400$ & 9:20 & Mobile Screen Resolution \\
    $1170 \times 2532$ & 90:195 & Mobile Screen Resolution \\
    $1440 \times 3088$ & 90:193 & Mobile Screen Resolution \\
    \bottomrule
    \end{tabular}
}
\end{center}
\label{table:resolutions}
\end{table}

\begin{table}[!t]
\caption{Each device used in our performance evaluations.}
\begin{center}
\resizebox{0.95\columnwidth}{!}{
  \begin{tabular}{c c c c}
    \toprule
    {\bf Device} & {\bf CPU Cores} & {\bf GPU} & {\bf Resolution} \\ 
    \midrule
    PC Workstation & 12 Cores & RTX 2080 Super (432 Cores) & $1920 \times 1080$ \\
    2021 Macbook Air & 8 Cores & Apple M1 (8 Cores) & $2560 \times 1600$ \\
    Samsung Galaxy S20 Ultra & 8 Cores & Mali G77 (11 Cores) & $3200 \times 1440$ \\
    iPhone 13 Pro & 6 Cores & A15 (5 Cores) & $2532 \times 1170$ \\
    \bottomrule
    \end{tabular}
}
\end{center}
\label{table:devices}
\end{table}

\begin{table}[t]
\caption{Total number of images/frames for each dataset.}
\begin{center}
\resizebox{\columnwidth}{!}{
  \begin{tabular}{c c c c c}
    \toprule
    {\bf DIV2K Train} & {\bf DIV2K Valid} & {\bf RICO Valid} & {\bf DAVIS 480p} & {\bf DAVIS 1080p} \\ 
    \midrule
    800 & 100 & 1460 & 761 & 761 \\
    \bottomrule
    \end{tabular}
}
\end{center}
\label{table:image_count}
\end{table}

In \cref{table:image_count}, we provide the distributions each dataset used in our evaluations in \cref{subsec:results_similarity,subsec:results_semantic}.

% In \cref{table:resolutions}, we provide a comprehensive list of each image resolution size  used in our performance evaluations in \cref{subsec:results_performance}. This table contains the resolution size, aspect ratio, and the purpose of including it in our performance evaluations.

Finally, we provide a comprehensive comparison of many other closely related works that seek to provide defenses against shoulder surfing. We do not include shoulder surfer detection systems, since they are dissimilar from \name. \Cref{table:relatedwork} demonstrates how \name is capable of protecting all of the information types except for PIN. Unlike privacy films, it is also able to protect against shoulder surfers directly behind the user.

% \section{Supplemental Experiments}

% In this section, we provide several supplemental experiments and data from evaluating \name's efficacy. \Cref{fig:ssim_original} contains the SSIM scores comparing the full-sized protected images to the original source images.

% \input{FiguresTex/fig_ssim_original}

\section{Supplemental Images, Examples, Prototypes}
\label{sec:examples}

In this section, we provide several supplemental figures portraying \name protected screens and images. Note that due to artifacts from downsampling and camera capture, the supplementary images and examples provimded in the paper
are not as clear as they would appear to an intended user viewing the original in-person. 

\Cref{fig:example_shouldersurfer} shows that without a camera with optical zoom, it is very difficult 
to discern any of the text on the protected screen compared to the unprotected screen. 
Information such as names, phone numbers, and account details are redacted.

\section{Equations}

\begin{equation} \label{eq:rms}
    \mathsf{rms} = \sqrt{\dfrac{x^{2} + y^{2}}{2}}
    \quad\text{and}\quad 
     y = \sqrt{\left(\mathsf{rms}^{2} \cdot 2\right) - x^{2}}.
    % Where x and y are RGB colors
\end{equation}

\begin{equation} \label{eq:ssim}
    SSIM\left(x, y\right) = \dfrac{\left(2\mu_{x}\mu_{y} + c_1\right)\left(2\sigma_{xy} + 
    c_2\right)}{\left(\mu_x^2 + \mu_y^2 + c_1\right)\left(\sigma_x^2 + \sigma_y^2 + c_2\right)}
\end{equation}

\begin{equation} \label{eq:angular_size}
    \delta = 2 \arctan \left(\dfrac{d}{2D} \right).
\end{equation}

\begin{figure}[t!]
\begin{subfigure}{0.48\columnwidth}
\centering
  \centering
  \includegraphics[width=\columnwidth]{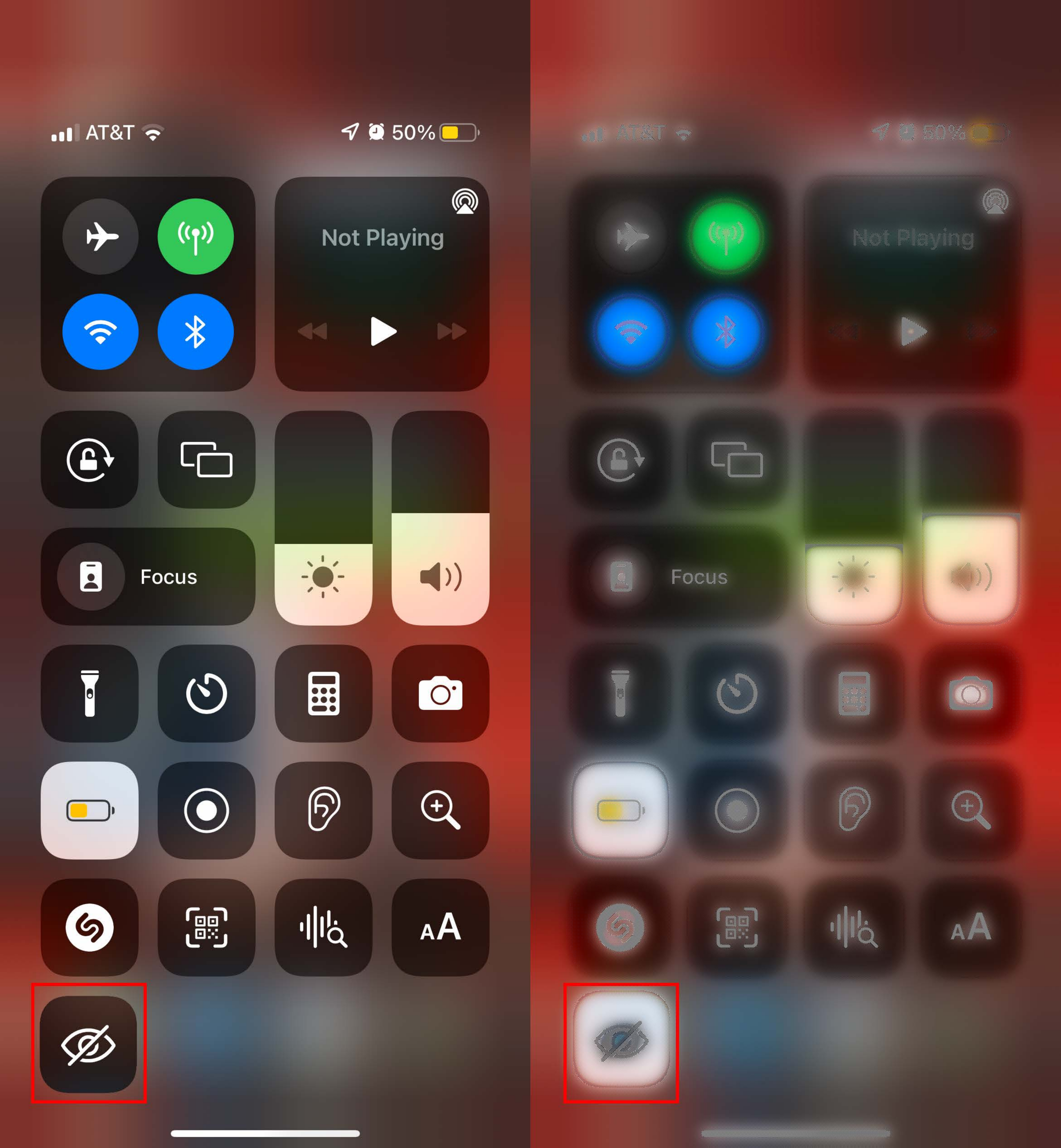}
  \vspace*{-0.2in}
\end{subfigure}
\begin{subfigure}{0.48\columnwidth}
  \centering
  \includegraphics[width=\columnwidth]{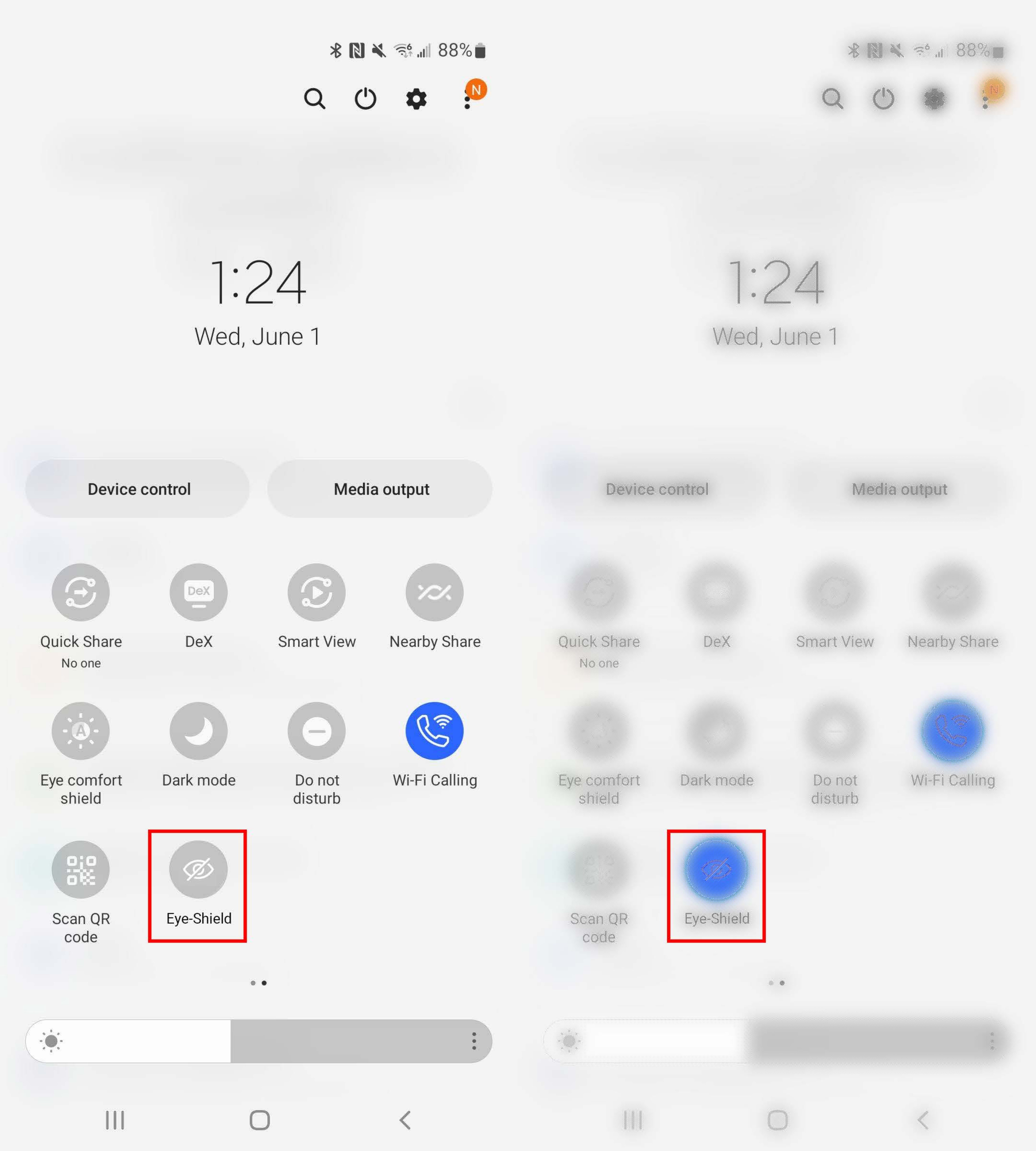}
  \vspace*{-0.2in}
\end{subfigure}
\caption{iOS (left) and Android (right) UI prototypes for toggling \name. The red box highlights the toggle.}
\label{fig:prototype}
\end{figure}

\begin{figure*}[t!]
\centering
\begin{subfigure}{0.4655\textwidth}
  \centering
  \includegraphics[width=\textwidth]{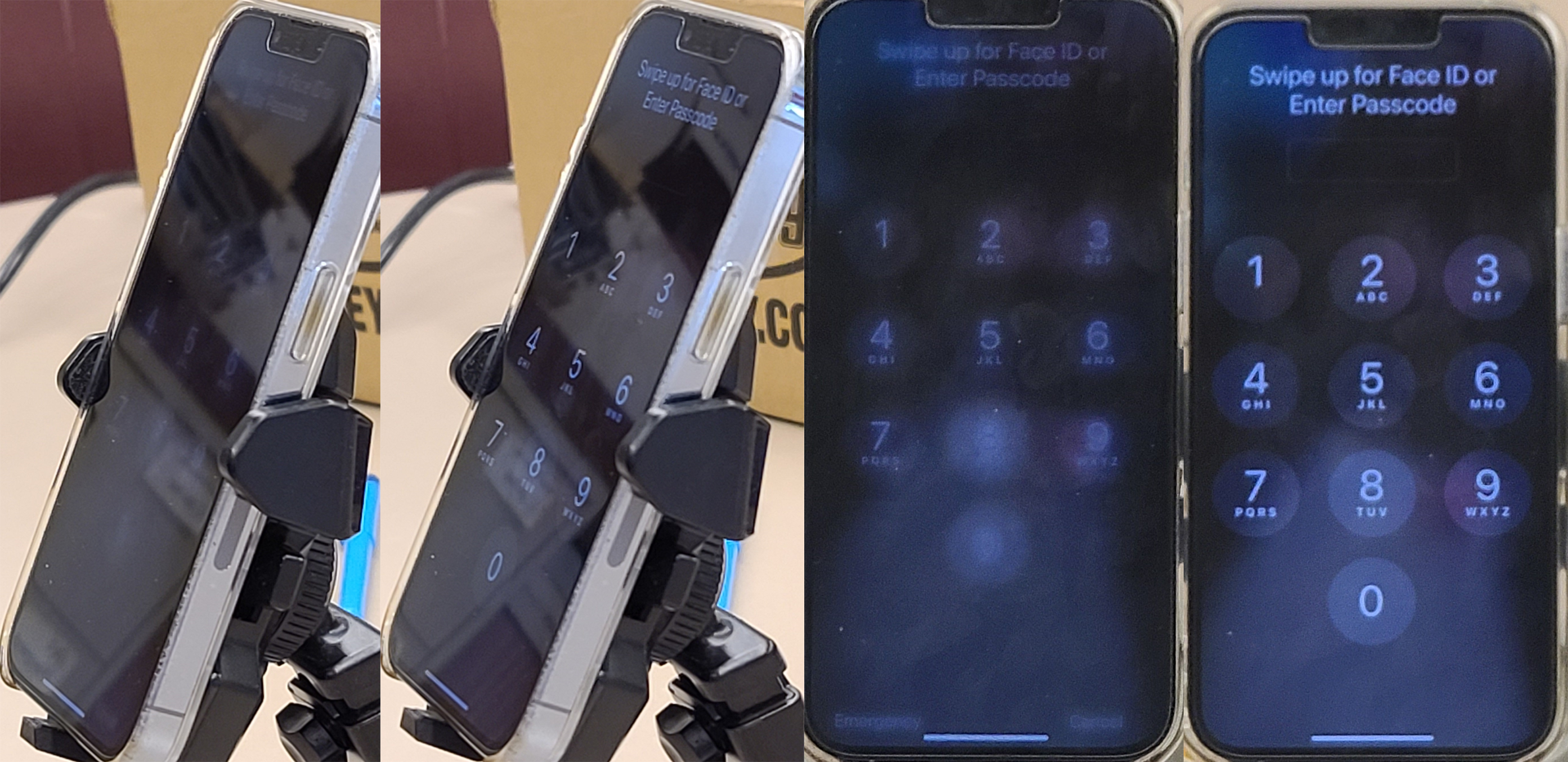}
  \caption{Distance of (left) \textit{19.7'' and 45\degree angle} and (right) \textit{41''}.}
  \label{fig:example_shouldersurfer}
\end{subfigure}
\begin{subfigure}{0.47\textwidth}
  \centering
  \includegraphics[width=\textwidth]{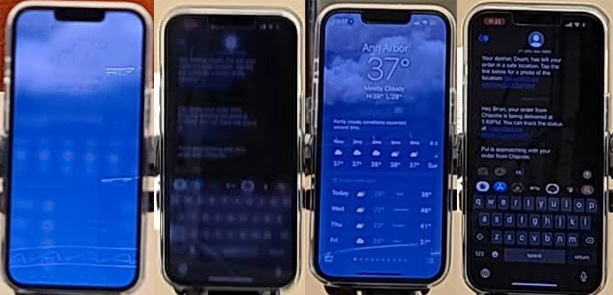}
  \caption{Distance of \textit{41''}.}
  \label{fig:example_shouldersurfer}
\end{subfigure}
  \caption{Photos of protected and unprotected screens captured with (a) \textit{48MP, f/1.8, 103mm smartphone camera at $5\times$ zoom.} and (b) \textit{108MP, f/1.8, 26mm smartphone camera at $1\times$ zoom.}}
  \label{fig:example_shouldersurfer}
\end{figure*}

% \begin{figure*}[h!]
%   \centering
%   \includegraphics[width=0.8\textwidth]{Figures/camera_examples/closeup_protected.pdf}
%   \caption{Photos of screens protected by \name (left) and the unprotected original screens (right). Photos were taken at a distance of \textit{10''}, from the perspective of the \textit{intended user}. The images were captured with a \textit{108MP, f/1.8, 26mm smartphone camera at $1\times$ zoom.}}
%   \label{fig:example_closeup}
% \end{figure*}

% \begin{figure*}[h!]
%   \centering
%   \includegraphics[width=\textwidth]{Figures/camera_examples/faraway_protected_zoom.pdf}
%   \caption{Photos of screens protected by \name (left) and the unprotected original screens (right). Photos were taken at a distance of \textit{41''}, from the perspective of a \textit{shoulder surfer}. The images were captured with a \textit{48MP, f/1.8, 103mm smartphone camera at $5\times$ zoom.}}
%   \label{fig:example_faraway_zoom}
% \end{figure*}

% \begin{figure*}[h!]
%   \centering
%   \includegraphics[width=\textwidth]{Figures/camera_examples/faraway_protected.pdf}
%   \caption{Photos of screens protected by \name (left) and the unprotected original screens (right). Photos were taken at a distance of \textit{41''}, from the perspective of a \textit{shoulder surfer}. The images were captured with a \textit{108MP, f/3.5, 26mm smartphone camera at $1\times$ zoom.}}
%   \label{fig:example_faraway}
% \end{figure*}

\begin{table*}[b]
\caption{A comparison of \name with other non-detection based shoulder surfing defense mechanisms.}
\begin{center}
\scalebox{0.7}{
  \begin{tabular}{c c c c c c c c}
    \toprule
    {} & {\bf \name} & {\bf \hidescreen~\cite{chen2019keep}} & {\bf Scrawl~\cite{eiband2016my}} & {\bf Gallery~\cite{von2016you}} & {\bf IllusionPIN~\cite{papadopoulos2017illusionpin}} & {\bf Privacy Shade~\cite{blackberry2022privacyshade}} & {\bf Privacy Film~\cite{3mprivacyfilm}} \\ 
    \midrule
    PIN & X & \checkmark & X & X & \checkmark & X & X \\
    Text & \checkmark & \checkmark & \checkmark & X & X & \checkmark & \checkmark \\
    Grayscale Photos & \checkmark & \checkmark & X & \checkmark & X & \checkmark & \checkmark \\
    Color Photos & \checkmark & X & X & \checkmark & X & \checkmark & \checkmark \\
    Videos & \checkmark & X & X & X & X & X & \checkmark \\
    Mobile UIs & \checkmark & X & X & X & X & \checkmark & \checkmark \\
    Entire Screen & \checkmark & X & X & X & X & X & \checkmark \\
    No Interruption & \checkmark & \checkmark & \checkmark & \checkmark & \checkmark & X & \checkmark \\
    Protects Behind & \checkmark & \checkmark & \checkmark & \checkmark & \checkmark & \checkmark & X \\
    \bottomrule
    \end{tabular}
}
\end{center}
\label{table:relatedwork}
\end{table*}

\begin{table*}[b]
\caption{Main discussion questions for interview. Exact phrasing differed between participants, but overall topics were consistent.}
\begin{center}
\scalebox{0.7}{
  \begin{tabular}{c l l}
    \toprule
    {\bf Code} & {\bf Question} & {Count/Responses} \\ 
    \midrule
    Q1 & Does previous familiarity with a particular app UI help with your use of Eye-Shield? & 3\\
    Q2 & Would you prefer using Eye-Shield over using a physical privacy film? & 7\\
    Q3 & Would you use Eye-Shield to protect financial app information or PIN entry? & 7\\
    Q4 & Would you use Eye-Shield to protect your text messages or photos? & 3\\
    Q5 & Did you find the blurring that Eye-Shield causes annoying as the intended user? & 7\\
    Q6 & Did you experience any eye strain when using Eye-Shield as the intended user? & 3\\
    Q7 & Would you feel uncomfortable if you saw someone shoulder-surfing your device? & 8\\
    Q8 & Have you experienced shoulder surfing before as the shoulder surfer or the victim? & 13\\
    Q9 & Do you prefer blurring over pixelation as the intended user? & 3\\
    Q10 & In what use cases could you see yourself using Eye-Shield? & Time-sensitive tasks, public transport, family snooping, lecture, church\\
    Q11 & What method would you prefer Eye-Shield to be activated with? & Widget in control center, blurring/brightness meter, automatic activation\\
    Q12 & What protection methods have you used to protect yourself from shoulder surfing? & Cover phone, lower brightness, stop usage, privacy film, Face ID\\
    Q13 & What methods did you find improved the usability as the intended user? & Zooming in, increasing brightness, leaning in\\
    Q14 & Various useful comments the participants brought up. & Difficulty with trifocals, accessibility for languages, pixelation eyestrain, \\
     &  & PIN-entry default, older people less usage, annoyance for reading quickly\\
    \bottomrule
    \end{tabular}
}
\end{center}
\label{table:qualitative}
\end{table*}

% \begin{figure*}[t!]
% \centering
% \begin{subfigure}{0.265\textwidth}
%   \centering
%   \includegraphics[width=\textwidth]{Figures/camera_examples/side_1_t_33.jpg}
%   \caption{Low Device Brightness}
%   \label{fig:darkest_screen}
% \end{subfigure}
% \begin{subfigure}{0.25\textwidth}
%   \centering
%   \includegraphics[width=\textwidth]{Figures/camera_examples/side_1_t_66.jpg}
%   \caption{Moderate Device Brightness}
%   \label{fig:moderate_screen}
% \end{subfigure}
% \begin{subfigure}{0.251\textwidth}
%   \centering
%   \includegraphics[width=\textwidth]{Figures/camera_examples/side_1_t_100.jpg}
%   \caption{High Device Brightness}
%   \label{fig:brightest_screen}
% \end{subfigure}
% \caption{Photos of protected images at various device screen brightness levels from the shoulder surfer's perspective.}
% \label{fig:device_brightness}
% \end{figure*}

% \begin{figure*}[t!]
%   \centering
%   \includegraphics[width=\textwidth]{Figures/examples/close_protected/contrast_example.JPG}
%   \caption{Examples of a protected mobile UI with varying contrast settings from normal contrast (left) to low contrast (right)}
%   \label{fig:contrast_example}
% \end{figure*}

% \input{FiguresTex/fig_entropy}

% Finally, we also provide a heatmap demonstrating the colors with the largest range of viable complementary colors, an important determination for computing the set of colors that are too high contrast to sufficiently protect using \name.

% \input{FiguresTex/fig_color_range}

% \input{FiguresTex/fig_examples_gallery}
% \input{Source/900_Revisions.tex}

The grid square size $g$ has a direct effect on the range at which the user can effectively operate their device, and it also has an effect on the range that a shoulder surfer can obtain information from viewing the user's device screen information. The grid square size should ideally be based on 3 factors: (1) the distance from the user to the device screen, (2) the device's pixel density (pixels per square inch, ppi), and (3) the font size. We found it optimal to set $g=1\times1$ in most of our experiments performed on the iPhone 13 Pro which has a $2532 \times 1170$ resolution with 460 ppi display, the user viewing the device from $10''$,  and the default application font size. Eye-Shield can adapt this to other configurations with different pixel densities or different text sizes by using this default $g=1\times1$ for 460 ppi and medium font size. We can calculate the optimal $g$ by scaling this proportionally to the device's ppi and font size, and rounding to the nearest pixel. For example, the original HideScreen paper used this equation $(d + 12'')/3333$ (where $d=$distance from device to intended user) to calculate the optimal $g$ in inches, which for 460 ppi and $d=10''$ comes out to $g=3\times3$ pixel squares. This was optimal for HideScreen which used larger font sizes.

Where
$g = $grid square size in pixels (variable to solve for)

$f = $fontsize in inches (verified with 0.25)

$d = $distance to screen in inches (10'')

$p = $pixel density in pixels per inch (460 PPI iPhone 13 Pro)

$\alpha = $human eye resolving power in degrees (0.0167 degrees)
\\\\
Using the equation of angular size: $\alpha = 2\arctan\dfrac{g/p}{2d}$

We get: $g = p \cdot 2d \cdot \tan(\dfrac{\alpha}{2})$

Scaling for font size: $g = p \cdot 2d \cdot \tan(\dfrac{\alpha}{2}) \cdot 4f$

Example: $g = 460 \cdot 20'' \cdot \tan(\dfrac{0.0167}{2}) \cdot 1.0$

Example: $g = 1.34$ pixels

% The advantage of the Eye-Shield algorithm over previous algorithms such as HideScreen is its parallelization and scalability. Eye-Shield works with any range of colors and can parallelize the steps of blurring and grid generation. The algorithm also allows for much of the image processing (blurring and matrix operations) to be performed in parallel or on a GPU. Where HideScreen relied on iterating through each pixel, Eye-Shield can perform operations on entire images at once using consumer-grade smartphones with GPUs.

Based on how much privacy protection the user wants, they can choose several protection strength settings. The examples we define in our prototype include: "full", "strong", "moderate", and "weak" protection. The contrast is adjusted using the OpenCV addWeighted() function on both the original image and the blurred target image. The privacy protection settings alter the blurring ($\sigma$) and contrast parameters as seen below:

\begin{itemize}[noitemsep,leftmargin=0.4cm,topsep=5pt]
    \item Full: $\sigma=24$, contrast = 80/127
	\item Strong: $\sigma=20$, contrast = 100/127
	\item Moderate: $\sigma=16$, contrast = 115/127
	\item Weak: $\sigma=8$, contrast = 127/127
\end{itemize}

$\sigma$ refers to the standard deviation of the Gaussian blur filter. The Gaussian filter convolves over an image with a specific window size with a Gaussian distribution. This means that a larger $\sigma$ value blurs/blends together a larger radius of pixels compared to a smaller $\sigma$ value. This contrast parameter decreases the contrast of the output by computing:
\\\\
\noindent $\alpha = float(131 \cdot (contrast + 127)) / (127 \cdot (131 - contrast))$
\\
$\gamma = 127 \cdot (1 - \alpha)$
\\
$output\_img = (img \cdot \alpha) + \gamma$

\end{document}